\documentclass[11pt]{article}
\usepackage{lastpage}
\usepackage{amsmath,amssymb}

\usepackage{setspace}
\usepackage{soul}
\usepackage{multirow}
\usepackage[margin=1in]{geometry}
\usepackage{graphicx}
\usepackage{placeins}
\usepackage{titlecaps}
\usepackage[dvipsnames]{xcolor}
\usepackage{caption}
\usepackage{abstract}
\usepackage[UKenglish]{babel}
\usepackage[super,sort&compress]{natbib}
\setcitestyle{comma}
\usepackage{textcomp}

\usepackage{sidecap}
\sidecaptionvpos{figure}{c}

\usepackage{xurl}
\usepackage[colorlinks = true]{hyperref}
\hypersetup{hidelinks}

\usepackage{titlesec}
\titleformat{\section}
{\large\bfseries}
{\thesection.}{0.5em}{}
\titleformat{\subsection}
{\small\bfseries\slshape}
{\thesection.}{0.5em}{}

\usepackage{titling}
\usepackage{fancyhdr}
\title{\Large \scshape Effects of a Nuclear Detonation in Low Earth Orbit}

\author{\normalfont Zachary Petersen$^{1}$, Hong Qin$^{1}$, Sarah Thiele$^{2}$, and S\'{e}bastien Philippe$^{1,*}$}

\date{\footnotesize
    $^1$ Department of Nuclear Engineering and Engineering Physics, University of Wisconsin--Madison, \\
    Madison, WI, USA\\
    $^2$ Department of Astrophysical Sciences, Princeton University, \\
    Princeton, NJ, USA\\
    \bigskip
    $^*$correspondence to sphilippe@wisc.edu\\[2ex]
}

\usepackage{etoolbox}
\usepackage{multibib}
\newcites{supp}{Supplementary references}

\begin{document}

\renewcommand{\abstractname}{}
\renewcommand{\absnamepos}{empty}

\maketitle

\begin{abstract}
\onehalfspacing
\noindent

\textbf{Abstract:} More than 13,000 satellites operate in low Earth orbit (LEO), and concern has grown that nuclear weapons could be used in space to target them. Beyond the initial X-ray burst, electrons released by radioactive fission products can become trapped in Earth's magnetic field, repeatedly exposing satellites as an artificial radiation belt. Here we couple prompt radiation, electron trapping and belt evolution to satellite failures, collisions and atmospheric drag. For a 1-Mt detonation at 500~km over the South Atlantic, repeated belt exposure disables approximately 98\% of modeled satellites within one week in our central case. Tripling the assumed radiation tolerance delays widespread failure by only days: approximately 97\% are still disabled within two weeks. The loss of maneuverability sharply increases collision risk, with approximately 20 collisions expected within two years. Drag removes many disabled spacecraft from lower altitudes, but replacement satellites could remain exposed to radiation and debris for years. These findings show how a single detonation could devastate orbital infrastructure and support efforts to prevent the deployment and use of nuclear weapons in space.

\end{abstract}

\FloatBarrier
\section*{Introduction}

Low Earth orbit (LEO) has become critical infrastructure, supporting civilian and military communications, weather forecasting, Earth observation and military reconnaissance. Lower launch costs and short signal latency have driven the deployment of large commercial constellations~\cite{bennett2023}. By January 2026, more than 13,000 satellites occupied LEO, with many concentrated near 500--600~km~\cite{kelso2026} (Supplementary Fig.~\ref{fig:leo_population}). Armed forces are also expanding their use of these orbits: the U.S. Department of Defense is deploying hundreds of satellites for missile warning, missile tracking and military communications~\cite{gao2026}.

This growing civilian and military reliance on LEO has made it a potential target. In 2024, the United States stated that Russia was developing a nuclear anti-satellite capability~\cite{whitehouse2024}. Russia denied plans to place nuclear weapons in space~\cite{ap2024putin} but vetoed a UN Security Council resolution reaffirming the Outer Space Treaty's prohibition on placing nuclear weapons in orbit~\cite{unsc2024}. These developments renewed concern that a nuclear weapon could be used against orbital infrastructure, with consequences extending beyond the satellites targeted. Such a weapon could be delivered by a missile or placed in orbit as a payload and detonated later.

The Starfish Prime nuclear test on 9 July 1962 provides a historical precedent for the widespread and lasting effects of a nuclear detonation in LEO. A 1.4-Mt device was detonated at approximately 400~km above Johnston Atoll in the Pacific. Electrons from radioactive fission-product decay became trapped in Earth's magnetic field, creating an artificial radiation belt that damaged satellites then in orbit or launched subsequently~\cite{hess1963,hoerlin1976,gombosi2017}. Unlike the prompt burst, a trapped belt repeatedly exposes spacecraft as they pass through it. A comparable event today would occur in a much more densely populated orbital environment.

Recent open assessments have focused primarily on prompt radiation. Fang and Mori\'c estimate that X-rays from a 1-Mt detonation would permanently disable at most about 5\% of non-military LEO satellites~\cite{fang2026}. The artificial belt can expose a much wider orbital population, but its formation depends on where the burst injects electrons and whether they remain trapped throughout their drift around Earth~\cite{cowee2012,cunningham2017}. Radiation failures also change the collision environment: disabled satellites can no longer maneuver or execute controlled disposal. Recent work shows how dependent crowded LEO is on continued collision avoidance~\cite{thiele2026}, while collision fragments can create additional hazards for large constellations~\cite{thiele2022}. Assessing the consequences of a detonation therefore requires linking radiation exposure to the subsequent evolution of intact spacecraft and debris.

Here, we develop an open, end-to-end framework that connects a nuclear detonation to radiation-driven satellite failures and the subsequent evolution of the orbital environment. We calculate prompt X-ray exposure and use an open-source burn-up code to obtain the electrons released by radioactive fission products~\cite{deTroullioud2021}. We follow the expansion of radioactive debris across Earth's magnetic field and identify the magnetic shells onto which electrons are released. Using a realistic geomagnetic field~\cite{beggan2026}, we determine which electrons remain trapped throughout their drift around Earth and which are lost to the atmosphere. We then evolve the artificial belt and calculate the radiation dose accumulated along individual satellite orbits. We adapt and couple existing collision and debris models~\cite{thiele2026,thiele2022}, adding explicit tracking of radiation-disabled spacecraft, object-dependent collision geometry, generation-specific masses and drag properties, mass-conserving fragmentation, breakup velocity kicks and orbital descent under changing solar activity. We first examine how detonation location determines whether an artificial radiation belt forms and which satellite orbits it crosses. We then follow satellite failures, collisions and orbital decay over a decade, and assess how the changing environment would affect a newly inserted satellite. The calculation provides a first-order assessment of the connected radiation and collision consequences, indicating that a single detonation could devastate LEO infrastructure and sustain orbital hazards for years.
\FloatBarrier
\section*{Results}

We develop a framework to follow a 1-Mt detonation from the initial radiation burst to satellite failures and their subsequent collision consequences. We assume that fission accounts for 50\% of the explosive yield, corresponding to approximately $7.25\times10^{25}$ fissions. We calculate how the radioactive debris expands through Earth's magnetic field, which magnetic shells it reaches, and whether the electrons it releases remain trapped as they drift around Earth. The magnetic field is evaluated for 14 January 2026, matching the epoch of the satellite catalogue used for the radiation calculation. We first compare two detonation locations chosen to maximize prompt exposure and artificial-belt effects, respectively. We then examine satellite losses and collisions following a burst at 500~km over $20^{\circ}$S, $35^{\circ}$W, near the South Atlantic Anomaly. This location places trapped electrons on compact magnetic shells that overlap densely populated LEO altitudes, concentrating the radiation and producing one of the most intense artificial belts in our location survey.

Radioactive debris ions release electrons through beta decay, so the electron source depends on how long the ions remain above the atmosphere. Observations of Starfish Prime debris arriving at the southern conjugate region, where the same magnetic field lines return toward Earth, guide our shorter release durations~\cite{dArcy1965}. We combine these durations with radioactive-decay histories calculated using an open-source burn-up code~\cite{deTroullioud2021}. We consider 0.40, 1.29 and 5.21 emitted electrons per fission, corresponding to release over approximately 2~s, 20~s and six months, respectively. The 20-s case is our central source; the 2-s case represents rapid atmospheric loss, while the six-month case provides an intentionally extreme upper-bound scenario, useful for assessing the consequences of prolonged debris retention rather than as a realistic expectation. The location-dependent trapping calculation then determines how many remain in the belt.

We follow the orbits of 13,300 satellite payloads, active in January 2026~\cite{kelso2026}, and calculate their radiation exposure behind 2~mm of aluminium. We consider two illustrative radiation-tolerance thresholds, motivated by an assumed fleet comprising 90\% civilian and 10\% military satellites. Combining assumed civilian and military tolerances of 10 and 300~krad(Si) gives a weighted average of approximately 40~krad(Si); combining 20 and 1,000~krad(Si) gives approximately 120~krad(Si). The military values correspond to component-level radiation-hardness assurance levels F and H in MIL-PRF-38535~\cite{milprf38535}. We apply each rounded threshold uniformly to the modeled population. The selected objects are concentrated near 500--600~km (Supplementary Fig.~\ref{fig:leo_population}). For the collision calculation, we combine the radiation-modeled objects with debris, rocket bodies and other payloads from the Space-Track catalogue for 14 January 2026, removing duplicate NORAD identifiers. Docked components are grouped as single collision targets.

Radiation-disabled spacecraft remain in orbit but are assumed to lose maneuverability. We follow collisions and breakup fragments in altitude bands, retaining mass and allowing fragments to move above and below their birth altitude after a velocity kick. Both intact objects and debris descend under atmospheric drag. The central calculation uses object-dependent geometric collision areas, with fixed-distance encounter areas retained as sensitivities. Equations, catalogue rules and assumptions are detailed in Methods and the Supplementary Materials.

\FloatBarrier
\subsection*{Burst location determines artificial belt formation}

\begin{figure}[!t]
    \centering
    \includegraphics[width=\linewidth]{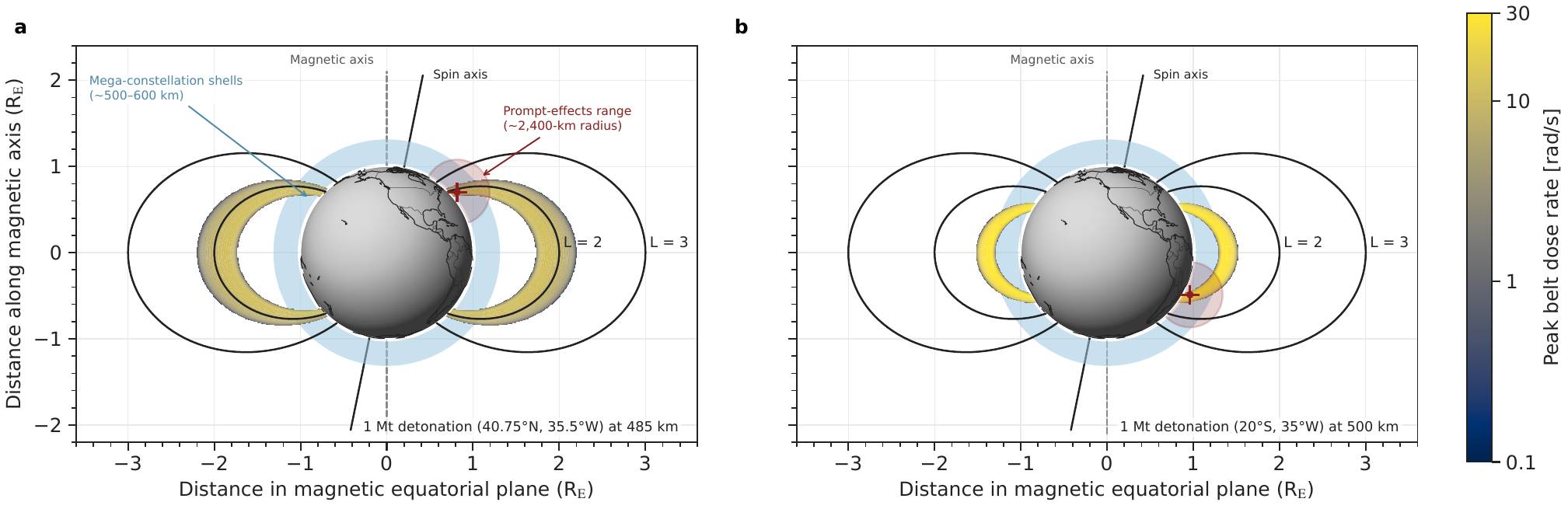}
    \caption{\textbf{Burst location changes the geometry and intensity of the artificial radiation belt.} Peak shell-average dose rate behind 2~mm of aluminium after a 1-Mt detonation, shown in the magnetic meridian of each burst for the central source of 1.29 emitted electrons per fission before trapping. \textbf{(a)} A burst at 485~km over $40.75^{\circ}$N, $35.5^{\circ}$W, selected near the highest instantaneous satellite density in the January catalogue to maximize prompt exposure. \textbf{(b)} A burst at 500~km over $20^{\circ}$S, $35^{\circ}$W, near the South Atlantic Anomaly, selected to produce an intense belt overlapping populated LEO altitudes. Red crosses mark the burst locations; red shading illustrates the approximately 2,400-km vacuum range within which prompt X-ray fluence exceeds the adopted 40~J\,m$^{-2}$ failure threshold~\cite{fang2026}. The blue band marks 200--2000-km altitudes, and black curves show $L=2$ and $L=3$. The display uses a centered-dipole projection; injection shells, trapping and belt intensities use IGRF-14 for 14 January 2026~\cite{beggan2026}. The burst in \textbf{(a)} populates shells $L=1.74$--2.19, with a peak dose rate of approximately 12~rad\,s$^{-1}$; the burst in \textbf{(b)} populates more compact shells, $L=1.31$--1.51, producing a peak of approximately 30~rad\,s$^{-1}$.}
    \label{fig:belt_geometry}
\end{figure}

We compare two detonation locations chosen to emphasize different effects (Fig.~\ref{fig:belt_geometry}). The first, at 485~km over $40.75^{\circ}$N, $35.5^{\circ}$W, places the burst near the highest instantaneous concentration of LEO satellites in the January catalogue to maximize prompt exposure. The second, at 500~km over $20^{\circ}$S, $35^{\circ}$W in the South Atlantic, favors an intense artificial radiation belt overlapping populated LEO altitudes. Both cases use the same 1-Mt yield and electron source.

Prompt X-rays and trapped electrons expose satellites in different ways. Prompt damage is limited to satellites with an unobstructed line of sight sufficiently close to the burst. The adopted fluence threshold of 40~J\,m$^{-2}$ corresponds to an unattenuated range of approximately 2,400~km. For the South Atlantic case, prompt X-rays disable 383 objects, or 2.9\% of the modeled fleet, at $t=0$. Subsequent exposure comes from repeated passages through the artificial radiation belt, allowing the effects to extend far beyond the region exposed to the initial burst.

\begin{figure}[!htbp]
    \centering
    \includegraphics[width=0.75\linewidth]{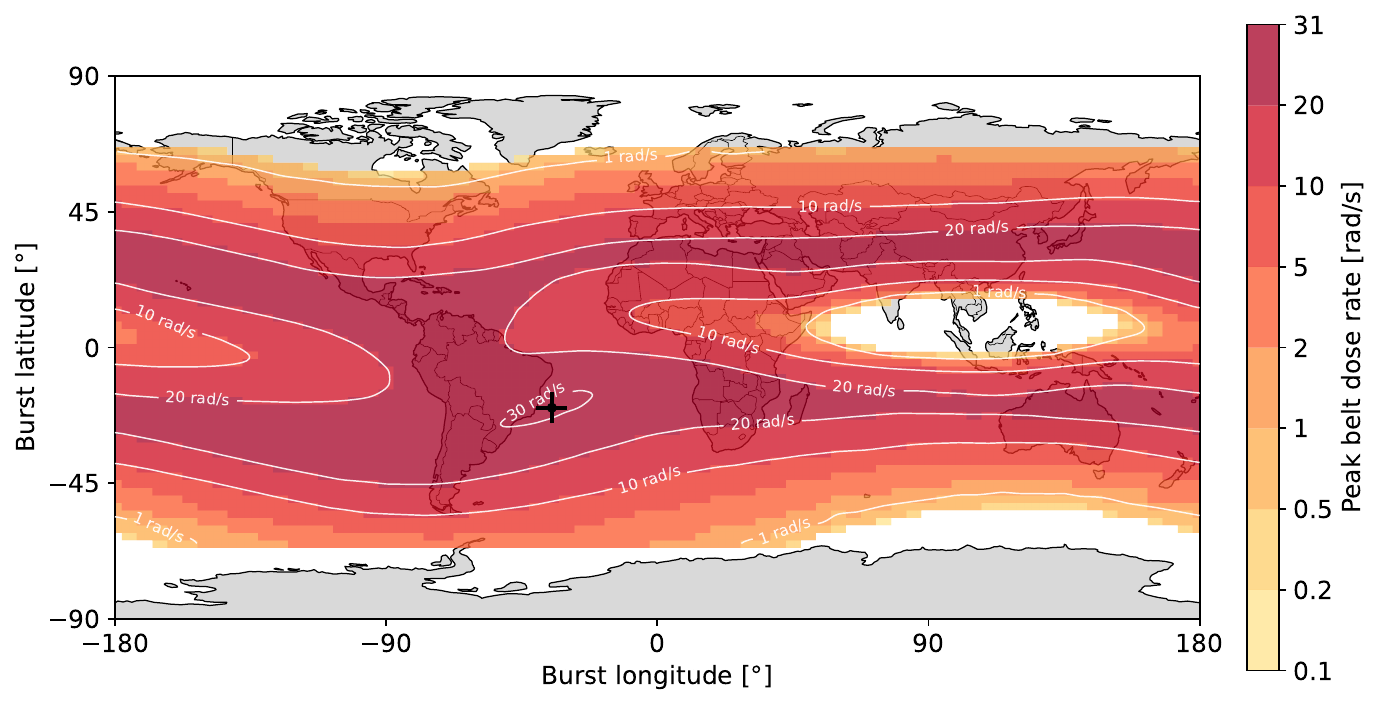}
    \caption{\textbf{Burst location determines whether electrons remain trapped and the intensity of the resulting belt.} Peak shell-average dose rate behind 2~mm of aluminium as a function of burst latitude and longitude for a 1-Mt detonation at 500~km, using the central source of 1.29 emitted electrons per fission and IGRF-14 for 14 January 2026~\cite{beggan2026}. White contours mark 1, 10, 20 and 30~rad\,s$^{-1}$; uncolored cells lie below the displayed 0.1~rad\,s$^{-1}$ limit or outside the calculation domain. For bursts near the magnetic equator over parts of the Indian Ocean and Indonesia, injected electrons reach shells whose mirror points descend below the 120-km atmospheric boundary over the South Atlantic Anomaly, preventing persistent trapping. This loss along the full drift path is accounted for through the drift loss cone~\cite{cladis1971,cunningham2017}. Injection over the South Atlantic can instead populate compact shells overlapping LEO, producing particularly intense belts. The cross marks the reference burst at $20^{\circ}$S, $35^{\circ}$W over the South Atlantic Anomaly.}

    \label{fig:location_map}
\end{figure}

Electron trapping depends on the full drift path around Earth, not only on the magnetic field near the detonation. An electron can be reflected above the atmosphere near its injection point but lost elsewhere along its path if its mirror point falls below the adopted atmospheric boundary of 120~km. Consequently, otherwise identical detonations at 500~km do not all produce a persistent trapped belt (Fig.~\ref{fig:location_map}). For bursts near the magnetic equator over parts of the Indian Ocean and Indonesia, opposite the South Atlantic Anomaly (SAA), injected electrons occupy shells whose mirror points descend into the atmosphere over the SAA. The weak-field region then acts as a loss point. Other locations support trapping but produce different belt intensities, with mapped peak dose rates varying by more than two orders of magnitude.

The two reference cases illustrate how belt geometry shapes exposure. The burst chosen to maximize prompt effects populates magnetic shells $L=1.74$--2.19 and retains approximately 79\% of emitted electrons. These shells cross the 500--600-km constellation altitudes at magnetic latitudes of about $38^{\circ}$--$45^{\circ}$ and produce a peak dose rate of approximately 12~rad\,s$^{-1}$ in the central source case (Fig.~\ref{fig:belt_geometry}a). The South Atlantic burst retains a smaller fraction, approximately 47\%, but places the electrons on more compact shells, $L=1.31$--1.51, crossing those altitudes at about $25^{\circ}$--$32^{\circ}$ magnetic latitude. Their smaller volumes concentrate the trapped electrons, increasing the peak dose rate to approximately 30~rad\,s$^{-1}$ (Fig.~\ref{fig:belt_geometry}b). The strongest mapped belts result from detonations above the South Atlantic Anomaly, where Earth's magnetic field is unusually weak (Fig.~\ref{fig:location_map}). Exposure therefore depends on both how many electrons remain trapped and how strongly they are concentrated on shells that intersect satellite orbits.

\FloatBarrier
\subsection*{Repeated belt exposure disables nearly the entire fleet}

\begin{figure}[!htbp]
    \centering
    \includegraphics[width=0.75\linewidth]{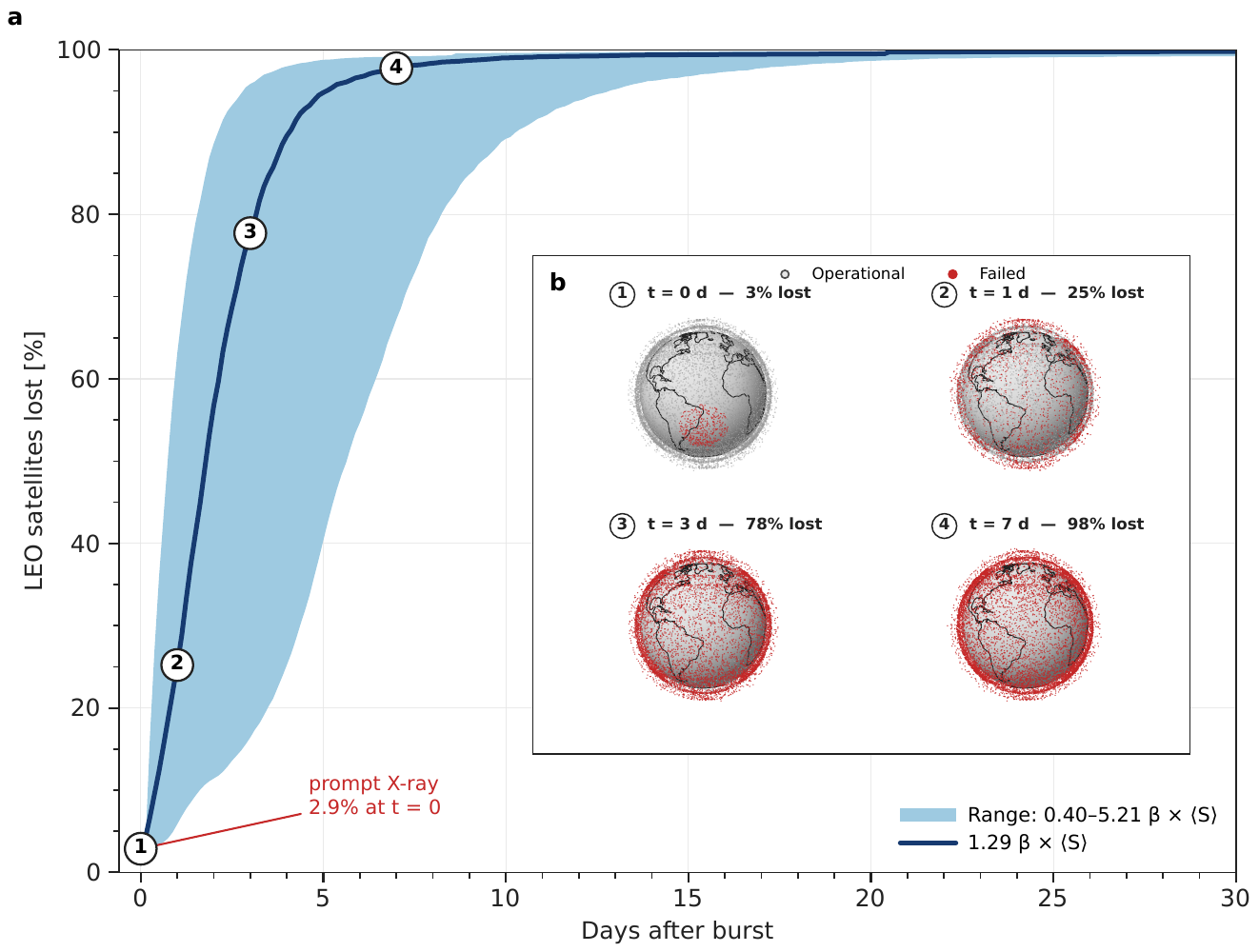}
\caption{\textbf{Repeated exposure to the artificial belt disables nearly the entire modeled fleet.} \textbf{(a)} Cumulative fraction of the 13,306-object fleet disabled after a 1-Mt detonation at 500~km over $20^{\circ}$S, $35^{\circ}$W, for a 40~krad(Si) failure threshold behind 2~mm of aluminium. The line denotes the central source of 1.29 emitted electrons per fission; shading spans the lower and upper sources of 0.40 and 5.21, each combined with the calculated trapping efficiency. Electron counts are derived from beta-activity histories calculated with ONIX~\cite{deTroullioud2021}. Prompt X-rays disable 2.9\% at $t=0$. Numbered markers identify the snapshots in \textbf{(b)}. \textbf{(b)} Fleet status at 0, 1, 3 and 7 days in the central source case, with satellite positions propagated to each snapshot time. Gray denotes operational objects and red denotes radiation-disabled objects. Losses rise from approximately 3\% initially to 25\%, 78\% and 98\%, respectively. Results for both tolerance thresholds and all three source cases are given in Supplementary Table~\ref{tab:radiation_loss_cases}.}
    \label{fig:fleet_loss}
\end{figure}

For the South Atlantic burst, repeated passages through the artificial radiation belt rapidly spread failures beyond the satellites disabled by prompt X-rays (Fig.~\ref{fig:fleet_loss}). With the lower radiation-tolerance threshold of 40~krad(Si) and the central electron source, approximately 25\% of the fleet is disabled after one day, 78\% after three days and 98\% after one week. Half the fleet crosses the failure threshold within 1.9 days and 90\% within 4.1 days. By day 15, 99.4\% is disabled. The snapshots in Fig.~\ref{fig:fleet_loss}b show this progression from localized prompt losses to failures throughout the orbital population.

The number of electrons supplied by the radioactive debris changes how quickly these failures occur. The upper source case reaches 90\% fleet loss in 2.1 days, compared with 10.4 days in the lower case. Nevertheless, all three source cases disable more than 99\% of the fleet within 30 days at the 40~krad(Si) threshold. Across these assumptions, repeated belt exposure therefore dominates the initial radiation losses.

Increasing the representative radiation-tolerance threshold to 120~krad(Si) delays failures (Supplementary Table~\ref{tab:radiation_loss_cases}). In the central source case, approximately 72\% of the fleet is disabled within one week, 97\% within 15 days and more than 99\% within 30 days. At this higher threshold, the upper and lower electron-source cases produce approximately 99.8\% and 90\% fleet loss by day 30, respectively. Greater radiation tolerance therefore extends survival, particularly when fewer electrons are released, but does not prevent widespread losses in any of the three source cases.

Lower yields delay failures but can still cause extensive fleet losses (Supplementary Table~\ref{tab:yield_loss}). With the central electron source, a 100-kt detonation at the same location and altitude disables 90\% of the fleet within approximately one month at 40~krad(Si) and three months at 120~krad(Si). After one year, losses reach approximately 99\% and 97\%, respectively. At 10~kt, radiation tolerance makes a larger difference: approximately 92\% of the fleet is disabled within one year at 40~krad(Si), compared with 19\% at 120~krad(Si).

\FloatBarrier
\subsection*{Loss of maneuverability produces an early collision surge}

Radiation-disabled spacecraft remain in orbit but are assumed unable to maneuver. Their loss of maneuverability increases collisions with other payloads, rocket bodies and debris. We report expected collision counts across all combinations of these objects, including debris--debris impacts and both catastrophic and cratering outcomes. These are population-level expectations, not individually simulated events. Debris participating in collisions has a characteristic length of at least 10~cm. We describe the changing hazard with an avoidance-modified CRASH Clock~\cite{thiele2026}, defined as the inverse instantaneous expected collision rate across the modeled population. We use this clock as an environmental indicator of the stress that a nuclear detonation places on LEO. It gives the mean waiting time to a collision anywhere in that population if the current rate were held constant, not the lifetime of an individual satellite.

\begin{figure}[!htbp]
\centering
\includegraphics[width=0.97\textwidth]{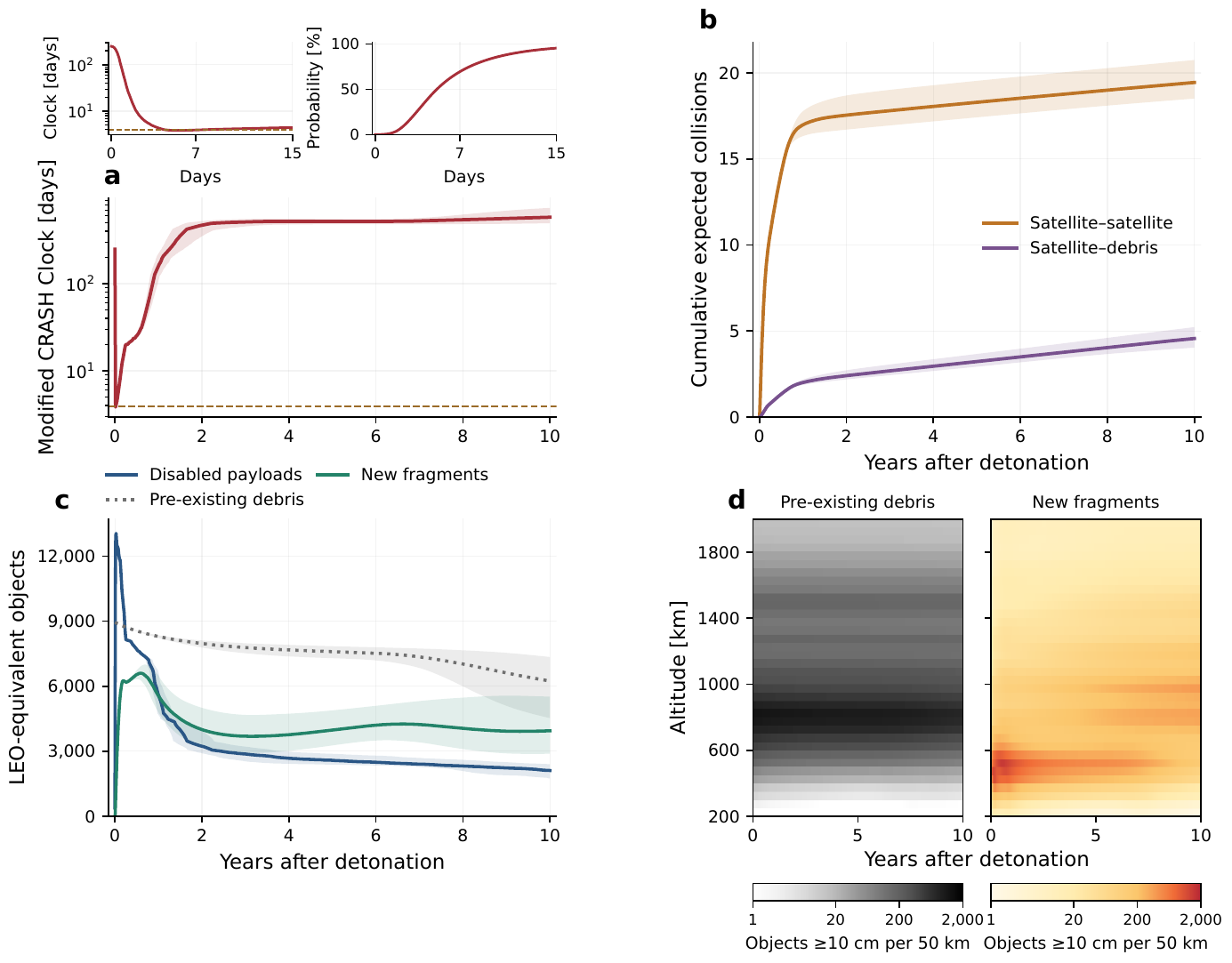}

\caption{\textbf{Radiation failures produce an early collision surge followed by altitude-dependent clearance.} Results for a 1-Mt detonation at 500~km over $20^{\circ}$S, $35^{\circ}$W, with the central electron source and a uniform 40~krad(Si) threshold behind 2~mm of aluminium. \textbf{(a)} Avoidance-modified global CRASH Clock, the inverse instantaneous expected collision rate across all modeled pairs, including catastrophic and cratering outcomes, rocket-body encounters and debris--debris impacts~\cite{thiele2026}. Small plots show the clock during the first 15 days and the cumulative probability of at least one collision. The dashed line marks the initial no-maneuvering reference calculated using the same catalogue and geometric cross-sections. \textbf{(b)} Cumulative expected satellite--satellite and satellite--debris collisions. Rocket-body and debris--debris encounters contribute to the reported totals but are omitted from this panel. \textbf{(c)} Intact radiation-disabled payloads, pre-existing catalogued debris and new collision fragments, weighted by the fraction of each orbit spent below 2000~km. Debris counts refer to characteristic lengths of at least 10~cm. Solid curves in \textbf{(a)--(c)} use the central solar-flux path; shading spans results from the NASA/MSFC 5th- and 95th-percentile paths, which alter atmospheric density and drag. These ranges show sensitivity to atmospheric forcing, not statistical confidence intervals. \textbf{(d)} Expected pre-existing debris and new fragments per 50-km altitude band, shown in grayscale and color, respectively. Both distributions use logarithmic scales spanning 1--2000 objects per band. New fragments receive breakup velocity kicks, and both debris populations move between altitudes under orbit-averaged drag~\cite{thiele2022,johnson2001}. Geometric cross-sections and masses depend on object class and Starlink generation. Atmospheric forcing follows the NASA/MSFC solar-flux paths and CIRA density profiles~\cite{nasa2026,cira2012}; model assumptions are detailed in Methods.}
\label{fig:combined_temporal_effects}
\end{figure}

In the central calculation, the clock falls from approximately 250 days immediately after the prompt burst to about 4 days within two weeks (Fig.~\ref{fig:combined_temporal_effects}a). A reference calculation in which all payloads lose maneuverability at the outset gives a clock of about 4 days for the same initial catalogue and geometric cross-sections. The separately reported January 2026 benchmark gives a similar value under different encounter assumptions~\cite{osi2026crash}.

Treating collisions as independent events occurring at the modeled rate, we estimate a 95\% probability of at least one collision within 15 days. By that time, the expected collision count is about 3, and approximately 2,400 new fragments at least 10~cm in size remain in the modeled LEO population. By day 30, about 6 collisions are expected in total, with approximately 4,400 new fragments remaining.

Approximately 20 collisions are expected within two years and 26 over a decade, compared with 12 over the decade in the matched no-detonation model. About 75\% of the detonation total occurs in the first year. The decade includes approximately 22 catastrophic collisions and 4 cratering events. Satellite--satellite collisions contribute about 19 events and satellite--debris collisions about 5 (Fig.~\ref{fig:combined_temporal_effects}b). Approximately 2 involve rocket bodies, while debris--debris impacts contribute fewer than 0.1 expected collision.

In the first year, approximately 19 collisions are expected after the detonation, compared with about one in the matched no-detonation calculation. Atmospheric drag subsequently removes many disabled spacecraft, reducing the additional collision hazard caused by the detonation.

\FloatBarrier
\subsection*{Drag removes lower-altitude spacecraft while hazards persist higher up}

Under central solar forcing, approximately 5,600 intact radiation-disabled payloads remain after one year, 3,200 after two years and 2,100 after ten years (Fig.~\ref{fig:combined_temporal_effects}c). These counts are weighted by the fraction of each object's orbit spent below 2000~km. As drag removes collision targets, the global CRASH Clock lengthens to approximately 470 days after two years and 580 days after a decade.

Drag removes many objects from lower altitudes, while spacecraft and debris higher up remain longer and continue to produce fragments through collisions. After ten years, approximately 3,900 new fragments and 6,200 pre-existing debris objects remain in the modeled LEO population. Breakup velocity kicks spread new fragments above and below their formation altitude, and atmospheric drag subsequently lowers their orbits (Fig.~\ref{fig:combined_temporal_effects}d).

Between years 2 and 10, approximately 10 additional collisions are expected in the no-detonation control, compared with about 6 after the detonation. This does not indicate a growing collision rate in the control: its rate remains nearly constant at approximately 1.2 collisions per year. Over the same interval, the detonation-case rate falls from approximately 0.8 to 0.6 collisions per year. Operational spacecraft maintain their orbits in the control, whereas many disabled spacecraft descend and reenter after the detonation. The lower late-time collision rate therefore reflects the removal of collision targets, not restoration of the orbital environment.

The main collision response persists across the tested solar-cycle and Starlink drag assumptions: an early increase follows the loss of maneuverability, then subsides as drag removes disabled spacecraft. The alternative solar paths give approximately 24--28 expected collisions over ten years, compared with 26 centrally; varying Starlink drag properties gives approximately 17--35 (Supplementary Table~\ref{tab:collision_sensitivity}). Collision-area assumptions produce a wider range, approximately 16--88. The larger fixed areas in the 10-5-10 prescription give global clocks of approximately 63 days at year 2 and 75 days at year 10, compared with 470 and 580 days using geometric cross-sections. These choices affect the magnitude of the collision response, while the initial fleet losses are driven by radiation.

Atmospheric removal is particularly important over the decade. Holding the low-activity density profile fixed gives approximately 477 collisions, compared with 26 under the central solar-cycle forcing. This deliberately sustained low-density case illustrates the consequences of weak removal rather than a forecast of atmospheric conditions. The sensitivity ranges are not statistical confidence intervals. Numerical checks change the two-year collision count by less than 0.5\%, substantially less than the tested physical assumptions.

To examine the environment for a replacement satellite, we calculate the collision rate for a test payload placed in a circular orbit at different altitudes and times after the detonation (Supplementary Fig.~\ref{fig:orbit_collision_rate_relative_to_background}). We represent it as a $9\times2.8\times0.2$-m spacecraft averaged over all orientations and consider two cases: no maneuvering and 99\% collision avoidance. At each altitude and time, we compare its instantaneous collision rate with that of an identical payload in the matched no-detonation environment.

At lower altitudes, the test satellite encounters fewer objects as drag clears disabled spacecraft and debris; its collision rate can eventually fall below the control because fewer targets remain. At higher altitudes, slower removal allows elevated rates to persist. These comparisons assume that the replacement satellite remains functional and exclude impacts from debris smaller than 10~cm. They therefore do not establish when operations could safely resume. Starfish Prime electrons persisted for several years~\cite{bostrom1970}, suggesting that replacement satellites could remain vulnerable to radiation even where collision activity has declined.

\FloatBarrier
\section*{Discussion}

An artificial radiation belt could turn a single nuclear detonation into a fleet-wide failure event. In the South Atlantic case, prompt X-rays account for only a small fraction of losses; repeated exposure to trapped electrons disables nearly the entire modeled fleet within days to weeks. Greater assumed radiation tolerance delays widespread failure by only days in the central electron-source case. Reducing the yield to 100~kt extends failures over weeks to months but still disables nearly the entire modeled fleet within one year in the central source case. At 10~kt, the outcome depends more strongly on radiation tolerance. The South Atlantic burst concentrates electrons on compact magnetic shells overlapping populated orbits, whereas some other locations lead to atmospheric loss rather than persistent trapping. Yield and altitude alone therefore do not determine the consequences.

Loss of maneuverability increases the modeled collision rate, with approximately 75\% of the decade's expected events occurring in the first year. Drag subsequently removes many lower-altitude targets, while objects higher up remain longer. This decline follows the removal of spacecraft and debris, not restoration of the lost services. Radiation may also outlast the initial surge: Starfish Prime electrons persisted for several years~\cite{bostrom1970}, so replacement satellites could remain vulnerable even after collision activity subsides.

Continued collisions among long-lived objects are expected to sustain debris production beyond the simulated decade, with the possibility of a slow buildup and eventual runaway growth on century timescales. A separate stability assessment identifies conditions for runaway growth across much of the 520--1000-km region in the March 2025 population~\cite{lewis2025}. Establishing the century-scale evolution after the modeled detonation requires longer integrations.

The main uncertainty in the electron source is how long radioactive debris ions remain above the atmosphere and continue to release electrons. We use three release durations: the shorter cases are guided by Starfish Prime observations of debris arrival in the conjugate atmosphere, while the longest provides an extreme upper-bound scenario. Together, these cases span a broad range of electron sources, yet all produce widespread satellite losses. We distribute the source over the magnetic shells reached by the expanding debris rather than following individual ions. Tracking their motion, trapping and atmospheric loss would better determine where and when electrons are released. More detailed shielding calculations and spacecraft-specific radiation tolerances would refine the timing and extent of failures, while dose estimates for newly launched satellites would help assess the duration of the hazard.

The collision model estimates rates from altitude-resolved populations rather than predicting individual encounters. Its geometric proxies and isotropic mixing simplify spacecraft shapes, attitudes and orbital geometry. The 99\% avoidance assumption is optimistic for newly generated or poorly tracked fragments, and impacts from debris below 10~cm are excluded. Masses and uncontrolled drag properties are estimates. Fragments receive velocity kicks and descend under orbit-averaged drag. Their initial orbits use a local circular-parent velocity approximation, and the model does not resolve inclination or orbital precession. Numerical tests check mass conservation and resolution but do not independently validate the physical assumptions. The ten-year horizon and exclusion of replenishment and sub-10-cm impacts limit the assessment of longer-term debris growth.

The effects could also reach the ground. A 1-Mt detonation at 500~km would produce an electromagnetic pulse (EMP) with a potential footprint spanning much of a continent~\cite{glasstone1977,gombosi2017,petersen2026}. A South Atlantic burst could therefore expose parts of South America to EMP while disrupting satellite services over a much wider area. Ground infrastructure damage requires a separate assessment, but is likely to be severe.

Atmospheric clearance also transfers spacecraft material to the atmosphere. During reentry, ablation releases metal vapors that condense into aerosol particles. Spacecraft reentry already supplies more lithium, aluminium, copper and lead than the natural cosmic-dust influx of these elements~\cite{murphy2023}. The reentry of thousands of disabled satellites would produce a substantial additional pulse of metals and could greatly enhance inputs above natural background levels for some elements. The resulting effects on stratospheric aerosol chemistry require a separate assessment.

Restoring orbital services would require more than replacing the lost spacecraft: new satellites would need to survive the remaining radiation and debris. The prospect of rapid infrastructure loss followed by years of disruption reinforces the need to prevent the deployment and use of nuclear weapons in space.

\FloatBarrier
\section*{Methods}

\subsection*{Model overview}

We developed \textit{open-hane} to calculate prompt radiation, fission-electron production, electron trapping and belt evolution, and the dose accumulated along satellite orbits. Radiation failures then change which spacecraft can maneuver in the coupled collision model, which follows fragmentation and atmospheric removal. Detailed population selection, magnetic-field, source, expansion, belt-evolution and collision-geometry calculations are given in Supplementary Methods.

Unless stated otherwise, the reference scenario is a 1-Mt detonation with a 50\% fission fraction at 500~km over $20^{\circ}$S, $35^{\circ}$W, on 14 January 2026 at 08:00 UTC. Satellites are assigned 2~mm of aluminium-equivalent shielding. The central radiation source is 1.29 emitted electrons per fission before trapping. Radiation histories cover one year; the coupled collision calculation extends to ten years.

\subsection*{Orbital population}

The radiation calculation propagates the 14 January 2026 CelesTrak elements with SGP4~\cite{kelso2026}, retaining 13,306 objects with successful propagation and instantaneous altitudes within 0--2000~km at the burst epoch. Radiation-loss fractions use this population. The collision inventory adds Space-Track payloads, rocket bodies and debris from the same date, deduplicates NORAD identifiers and groups docked components into single targets. Objects without radiation histories receive no avoidance credit. Counts are weighted by the fraction of each orbit spent below 2000~km. Neither the detonation nor control calculation includes launches or scheduled disposal. Full selection rules and population counts are given in Supplementary Methods.

\subsection*{Geomagnetic field}

We use IGRF-14 for 14 January 2026~\cite{beggan2026}, tracing field lines in both directions. Magnetic shells are labelled by their minimum field strength, $L=(B_0/B_{\min})^{1/3}$ with $B_0=31,200$~nT. Electrons are lost when their mirror points fall below 120~km anywhere along their drift around Earth. Field representation, tracing and lookup grids are detailed in Supplementary Methods.

\subsection*{Prompt X-rays}

The prompt-effects calculation follows Fang and Mori\'c~\cite{fang2026}, extended here to include energy-dependent atmospheric attenuation. We assume that X-rays carry a fraction $f_x=0.7$ of the explosive yield~\cite{glasstone1977}, with a blackbody spectrum at $kT=1$~keV normalized over 0.1--50~keV. The attenuated fluence at slant range $R$ is
\begin{equation}
\Phi=\frac{Y_{\mathrm{kt}}f_xE_{\mathrm{kt}}}{4\pi R^2}
\int p(E)\exp[-\mu_{\mathrm{air}}(E)\Sigma_{\mathrm{air}}]\,dE,
\label{eq:fluence}
\end{equation}
where $E_{\mathrm{kt}}=4.184\times10^{12}$~J\,kt$^{-1}$, $p(E)$ is the normalized blackbody energy spectrum, $\mu_{\mathrm{air}}$ is the mass attenuation coefficient of air~\cite{wang2025a}, and $\Sigma_{\mathrm{air}}$ is the atmospheric column in g\,cm$^{-2}$ integrated along the slant path~\cite{zhu2024}.

The column is integrated along each line of sight using a tabulated neutral-density profile approximating the U.S. Standard Atmosphere~\cite{noaa1976}, with nominal 5-km sampling and an atmospheric upper cutoff of 1000~km. Lines of sight intersecting the solid Earth are occulted. Prompt dose in silicon additionally accounts for aluminium attenuation and silicon mass energy absorption. This neutral-density profile is separate from both the ionized-air profile used for debris expansion and the CIRA profiles used for orbital drag.

Following Fang and Mori\'c~\cite{fang2026}, objects receiving $\Phi\geq40$~J\,m$^{-2}$ are counted as permanently disabled. This fluence criterion is applied independently of accumulated ionizing dose. Prompt neutrons, gamma rays and reversible upsets are not included in the coupled failure population.

\subsection*{Fission-electron source term}

We calculate the number of fissions from the yield, a 50\% fission fraction and 180~MeV per fission, giving approximately $7.25\times10^{25}$ fissions for 1~Mt. Electron production follows U-235 fission-product beta-activity tables calculated with ONIX~\cite{deTroullioud2021} and ENDF/B-VIII.0 nuclear data~\cite{brown2018}. Release over 2~s, 20~s and six months supplies 0.40, 1.29 and 5.21 electrons per fission, respectively. The shorter durations are guided by Starfish Prime debris arrival in the conjugate atmosphere~\cite{dArcy1965}; six months is an extreme upper bound. The short sources are injected initially; the prolonged source is added over time. Electron energies follow the Carter spectrum~\cite{carter1959}. Activity interpolation, source integration and spectral discretization are given in Supplementary Methods.

\subsection*{Burst region and injection shells}

We determine the source region from the magnetic-confinement and equal-mass arguments of the Los Alamos Electron Source Model~\cite{cowee2012}. Cross-field expansion stops at the smaller of the magnetic-confinement radius and the distance over which debris sweeps up its own mass in ionized air. Upward, downward and horizontal extents differ with atmospheric density. IGRF field lines through area-weighted source cells identify the injection shells; field-aligned expansion stops at the equal-mass limit or atmosphere. The reference burst reaches $L=1.31$--1.51. The expansion equations, density profile and discretization are given in Supplementary Methods.

\subsection*{Trapping}

An electron emitted at position $\mathbf{x}$ conserves its first adiabatic invariant and mirrors where $B_m=B(\mathbf{x})/\sin^2\alpha$, with $\alpha$ the local pitch angle~\cite{cladis1971}. It remains trapped only if its mirror points stay above the 120-km atmospheric boundary throughout its longitudinal drift. For a shell labelled by $B_{\min}$, the critical field is
\begin{equation}
B_c(B_{\min})=\min_{\phi}B\!\left(h=120~\mathrm{km};B_{\min},\phi\right).
\label{eq:criticalfield}
\end{equation}
For isotropic emission, the locally retained fraction is
\begin{equation}
S(\mathbf{x})=
\begin{cases}
\sqrt{1-B(\mathbf{x})/B_c}, & B(\mathbf{x})<B_c,\\
0, & B(\mathbf{x})\geq B_c.
\end{cases}
\label{eq:trapping}
\end{equation}
This includes the drift loss cone: electrons that mirror above the atmosphere locally may still be lost when their drift path reaches the weaker field over the South Atlantic.

Each cell's radioactive debris is treated as distributed uniformly in volume over its complete drift shell. The retained fraction $\bar S_c$ is averaged over the shell with volume weights proportional to $ds/B$. The total trapping efficiency and shell injection weights satisfy
\begin{equation}
\langle S\rangle=\sum_c w_c\bar S_c,
\qquad
\sum_i W_i=\langle S\rangle,
\label{eq:trappingweights}
\end{equation}
where $w_c$ is the cell's debris share and $W_i$ places its retained contribution on the belt grid. Contributions are shared between neighboring grid points and smoothed with a Gaussian width $\sigma_L=0.005$ for the reference calculation and 0.01 for the location map, followed by renormalization.

The injected population in energy bin $e$ and shell $i$ is $N_{e,i}=N_{\mathrm{fis}}n_{\beta}w_eW_i$, with subsequent source increments treated in the same way. No additional trapping factor is applied. For the reference burst, $\langle S\rangle=0.474$. Full-shell debris mixing and isotropic emission are model assumptions; wave-driven pitch-angle scattering during injection is not explicitly calculated.

\subsection*{Artificial-belt evolution}

The trapped population evolves by radial diffusion and parameterized electron losses, with continued injection in the prolonged-source case. We use the Brautigam--Albert diffusion coefficient~\cite{brautigam2000} at fixed $K_p=2$, zero-flux boundaries and three-hour timesteps. Prescribed energy- and shell-dependent lifetimes range from 0.5 days to eight years, representing atmospheric loss, long inner-belt retention and shorter slot-region lifetimes. Electron density uses dipole shell volumes above 80~km, separately from the 120-km IGRF trapping boundary. This is a reduced-order belt calculation~\cite{reeves2012,lee2024}; governing equations, coefficients and numerical grids are given in Supplementary Methods.

\subsection*{Electron dose and satellite failure}

The shell-average omnidirectional electron flux is $\phi_e=(N_{e,i}/V_i)v_e$. The dose rate behind the shield is
\begin{equation}
\dot D_i=\sum_e\frac{N_{e,i}}{V_i}\frac{v_e}{10^4}\,\mathcal R(E_e),
\label{eq:doserate}
\end{equation}
where $N/V$ is in m$^{-3}$, $v$ in m\,s$^{-1}$, and the factor $10^4$ converts flux to electrons cm$^{-2}$\,s$^{-1}$. For the central instantaneous source, the peak shell-average dose rate reported in Figs.~\ref{fig:belt_geometry} and~\ref{fig:location_map} is $\max_i\dot D_i(0)$. The response $\mathcal R$ is in rad(Si) per electron cm$^{-2}$:
\begin{equation}
\mathcal R(E)=1.602\times10^{-8}S_{\mathrm{Si}}(E)T_{\mathrm{Al}}(E)
+\mathcal R_{\mathrm{brem}}(E).
\label{eq:electronresponse}
\end{equation}

The 2-mm aluminium shield corresponds to an areal density of 0.54~g\,cm$^{-2}$. Silicon collision stopping powers and aluminium continuous-slowing-down ranges are taken from tabulated NIST ESTAR values~\cite{nistestar}. Transmission is approximated from electron range through the shield, with an additional cutoff when shield areal density exceeds the electron range. A small parameterized bremsstrahlung contribution is included.

Each object is propagated with SGP4 every 20 minutes for one year. TEME positions are rotated to Earth-fixed coordinates using Greenwich mean sidereal time. The shell label at each position is interpolated from an IGRF lookup grid with $2^{\circ}$ latitude, $4^{\circ}$ longitude and 100-km altitude spacing. The local dose rate is obtained from the shell average using
\begin{equation}
\dot D_s(t)=\dot D\!\left[L_s(t),t\right]
\frac{S[\mathbf{x}_s(t)]}{\bar S[L_s(t)]},
\label{eq:localdose}
\end{equation}
where $\bar S(L)$ is the volume-averaged retained fraction on that shell. This represents the assumed spatial distribution of electrons outside the loss cone.

Delayed dose is integrated along each orbit and added to prompt dose. Failure occurs when prompt fluence reaches 40~J\,m$^{-2}$ or accumulated prompt-plus-electron dose reaches the assigned uniform threshold of 40 or 120~krad(Si). These are illustrative tolerances, not measured spacecraft specifications or a literal civilian--military mixture. Failure is irreversible and resolved at the three-hour output cadence.

\subsection*{Collision rates and intact-object accounting}

We couple radiation failures to an altitude-resolved population model adapted from the CRASH Clock formalism and NASA breakup model~\cite{thiele2026,thiele2022,johnson2001}. Operational and disabled payloads, other payloads, rocket bodies, pre-existing debris and new fragments remain distinct. The central case uses 1.29 emitted electrons per fission and a uniform 40~krad(Si) threshold. Expected collision counts are obtained by integrating rates over time, including all pair types and catastrophic and cratering outcomes.

For unordered populations $a,b$ in a 10-km altitude shell $h$,
\begin{equation}
R_{ab,h}=s_{ab}p_{ab}\frac{v_h}{V_h}\mathcal{G}_{ab,h},
\qquad
v_h=\frac{4}{3}\sqrt{\frac{\mu_\oplus}{R_E+\bar h}},
\label{eq:pairrate}
\end{equation}
where $V_h$ is shell volume, $s_{ab}=1/2$ for identical populations and one otherwise, and $\mathcal{G}_{ab,h}$ combines population counts and geometric cross-sections, excluding self-encounters of intact objects. Relative velocities assume isotropic directions at the local circular speed. The residual-risk factor is $p_{ab}=0.01$ when either partner is operational and one otherwise, giving a 99\% avoidance benefit once per pair. Object accounting and the population-weighted cross-sections are detailed in Supplementary Methods.

\subsection*{Object-dependent collision geometry}

We calculate collision cross-sections from geometric proxies averaged over independently isotropic orientations, separately from aerodynamic areas. Starlink v1.x and V2 families use bus-and-array envelopes of $9\times2.8\times0.2$~m and $29\times4.1\times0.3$~m~\cite{spacex2022geometry}; other objects use GCAT dimensions~\cite{gcat2026}. The mean pair area follows the projected collision-region construction for extended objects~\cite{garciapelayo2018}. Station components, long antennas, tethers and missing dimensions receive the treatments detailed in Supplementary Methods.

For comparison, we use the encounter distances of Thiele and colleagues~\cite{thiele2026}: 10~m for intact--intact, 5~m for intact--debris and 10~cm for debris--debris pairs (10-5-10), and their smaller 4.8-m, 2.4-m and 10-cm prescription. These describe close encounters rather than object-dependent impacts. Sensitivity runs treat each modeled encounter as a collision and use the assigned areas for rates and fragmentation. Initial-rate checks against the published CRASH Clocks are given in Supplementary Methods.

\subsection*{Collision clock and probability}

The modified global CRASH Clock is the inverse instantaneous expected rate across all modeled pairs,
\begin{equation}
T_{\mathrm{all}}(t)=\left[\sum_{h,a,b}R_{ab,h}(t)\right]^{-1}.
\label{eq:collisionclock}
\end{equation}
It includes catastrophic and cratering events, rocket bodies and debris--debris pairs. It describes the waiting time to an event anywhere in the population if the current rate were held constant, not an individual satellite's lifetime. The probability diagnostic assumes a nonhomogeneous Poisson process,
\begin{equation}
P(\geq1;t)=1-\exp\!\left[-\int_0^tR_{\mathrm{all}}(t')\,dt'\right].
\label{eq:poisson}
\end{equation}
The plotted no-maneuvering reference uses the same initial population and geometric areas with avoidance switched off. The separately reported January CRASH Clock~\cite{osi2026crash} uses different encounter assumptions and is contextual, not a calibration target.

\subsection*{Intact masses, areas and atmospheric drag}

Starlink masses and uncontrolled area-to-mass ratios vary by generation; other intact objects use GCAT masses where available~\cite{gcat2026,sutton2025,spacex2024}. Adopted properties are given in Supplementary Table~\ref{tab:intact_properties}. Non-Starlink payloads use $A/m=0.010$~m$^2$\,kg$^{-1}$ and rocket bodies an assumed 0.005~m$^2$\,kg$^{-1}$. We adopt $C_D=2.2$ following ESA guidance~\cite{esa2025}. Operational payloads maintain their orbits until radiation failure; other objects undergo passive drag from the start. Orbit-averaged drag evolves perigee and eccentricity, updating residence in each altitude band~\cite{fitzpatrick}. Objects are removed when perigee reaches 200~km. Property assignments, mass changes and coordinate conventions are detailed in Supplementary Methods.

\subsection*{Fragmentation and orbital transport}

Collisions produce fragments according to the NASA Standard Breakup Model~\cite{johnson2001,thiele2022}. At a representative impact speed of 10~km\,s$^{-1}$, collisions are catastrophic when projectile energy per unit target mass reaches 40~kJ\,kg$^{-1}$; otherwise they erode the target. Fragment production is limited by available ejected mass. Material below 10~cm is retained in the mass budget but excluded from collision rates.

New fragments receive isotropic velocity kicks, added to a local circular-orbit velocity, which spread their orbits above and below the collision altitude. Pre-existing debris starts on its catalogued orbit without an additional kick. Both populations contribute to altitude bands according to orbital residence and descend under drag until perigee reaches 200~km. The model follows perigee and eccentricity without resolving inclination or orbital precession. Breakup equations, mass conservation, kick sampling and orbital transport are detailed in Supplementary Methods.

\subsection*{Atmospheric forcing and numerical integration}

Atmospheric density follows the low-, moderate- and high-activity CIRA-2012 profiles, interpolated using the NASA/MSFC October 2026 solar-flux history and forecast~\cite{cira2012,nasa2026}. The central case uses the median forecast; sensitivity cases use alternative solar paths, Starlink drag properties, fixed density profiles and encounter areas (Supplementary Table~\ref{tab:collision_sensitivity}).

The calculation advances in three-hour steps through day 60 and daily steps thereafter. Radiation-failure states are fixed after day 60, and dose is not recomputed as objects descend. Numerical checks and atmospheric interpolation are described in Supplementary Methods.

\subsection*{Collision hazard for an inserted satellite}

We calculate the instantaneous collision rate for a test payload on a circular orbit at different altitudes and times, using the same geometry and environment in detonation and no-detonation cases (Supplementary Fig.~\ref{fig:orbit_collision_rate_relative_to_background}). The test payload has a $9\times2.8\times0.2$-m geometric proxy, averaged over orientations, with either no maneuvering or 99\% collision avoidance. It does not alter the population and is assumed to survive radiation exposure. Impacts from debris below 10~cm are excluded. The rate calculation is given in Supplementary Methods.

\FloatBarrier
\subsection*{Acknowledgments}
Preliminary results were presented at the workshop on nuclear war effects modeling held at the University of Wisconsin--Madison on 2--5 June 2026 and at the 67th Annual Meeting of the Institute of Nuclear Materials Management on 3 August 2026~\cite{petersen2026}. Updated results were presented to the United Nations Independent Scientific Panel on the Effects of Nuclear War on 17 September 2026. The authors thank T.~Lockert for his contributions to the atmospheric attenuation of prompt X-rays and the conference version of this work, the workshop participants for their feedback, and L.~Revell for suggesting during the UN Panel presentation that we examine the collision consequences of radiation-induced satellite failures. We thank Z.~Fang and I.~Mori\'c for sharing their manuscript focused on prompt effects before publication. This work was supported by grants from the Andrew Carnegie Foundation (formerly the Carnegie Corporation of New York), Pax Sapiens and Longview Philanthropy to the University of Wisconsin--Madison (PI: S.~Philippe).
\FloatBarrier
\subsection*{Code availability}

The \textit{open-hane} source code will be available at \url{https://github.com/evident-lab/open-hane}. The collision and fragmentation modules are adapted from conjunctionSim/SatEvol (\url{https://github.com/norabolig/conjunctionSim}) and JunkySpace (\url{https://github.com/sarahthiele/junkyspace})~\cite{thiele2026,thiele2022}.

\FloatBarrier
\subsection*{Author contributions}
S.P. conceived and supervised the study and acquired funding. Z.P. developed \textit{open-hane}, including the prompt X-ray and artificial radiation-belt models, and performed the radiation simulations. S.P. adapted and coupled the collision, fragmentation and atmospheric-drag calculations to \textit{open-hane}, with input from S.T. H.Q. advised on electron trapping and radiation-belt physics. Z.P. and S.P. prepared the figures and wrote the original draft. S.P. revised and edited the manuscript. All authors reviewed the manuscript.

\FloatBarrier
\subsection*{Use of AI tools}
The authors used Claude Opus 5.5 (Anthropic) and ChatGPT and Codex with GPT-5.6 and GPT-6.1 (OpenAI) to assist with code development and review and manuscript editing. The authors reviewed the AI-assisted work and take responsibility for the calculations, methods, results and conclusions.

\FloatBarrier
\subsection*{Competing interests}
The authors declare no competing financial interests. S.P. is a member of the United Nations Independent Scientific Panel on the Effects of Nuclear War; this work was conducted independently of the Panel, and the views expressed are those of the authors. The funders had no role in the study design, analysis, interpretation of results or the decision to publish.

\bibliographystyle{unsrtnat}
\footnotesize\singlespacing\bibliography{references}

\begin{thebibliography}{44}
\providecommand{\natexlab}[1]{#1}
\providecommand{\url}[1]{\texttt{#1}}
\expandafter\ifx\csname urlstyle\endcsname\relax
  \providecommand{\doi}[1]{doi: #1}\else
  \providecommand{\doi}{doi: \begingroup \urlstyle{rm}\Url}\fi

\bibitem[Bennett and Kramer(2023)]{bennett2023}
Michael Bennett and Corinne Kramer.
\newblock {Large Constellations of Low-Altitude Satellites: A Primer}.
\newblock Technical report, Congressional Budget Office, Washington, DC, May
  2023.
\newblock URL \url{https://www.cbo.gov/publication/59175}.

\bibitem[Kelso(2026)]{kelso2026}
Thomas~Sean Kelso.
\newblock {{CelesTrak}: Current {NORAD} Two-Line Element Sets}, 2026.
\newblock URL \url{https://celestrak.org/NORAD/elements/}.

\bibitem[{U.S. Government Accountability Office}(2026)]{gao2026}
{U.S. Government Accountability Office}.
\newblock {Missile Warning Satellites: Space Development Agency Should Be More
  Realistic and Transparent About Risks to Capability Delivery}.
\newblock Technical Report GAO-26-107085, GAO, 2026.
\newblock URL \url{https://www.gao.gov/products/gao-26-107085}.

\bibitem[Sullivan(2024)]{whitehouse2024}
Jake Sullivan.
\newblock {Statement from National Security Advisor Jake Sullivan on Russia's
  Veto of the UN Security Council Resolution on the Outer Space Treaty}.
\newblock The White House, 24 April, 2024.
\newblock URL
  \url{https://bidenwhitehouse.archives.gov/briefing-room/statements-releases/2024/04/24/statement-from-national-security-advisor-jake-sullivan-on-russias-veto-of-the-un-security-council-resolution-on-the-outer-space-treaty/}.

\bibitem[{Associated Press}(2024)]{ap2024putin}
{Associated Press}.
\newblock {Putin says Russia has no intention of putting nuclear weapons in
  space, denying {US} claims}.
\newblock 20 February, 2024.
\newblock URL
  \url{https://apnews.com/article/15c91f8387ae3e6ec024e996755e321f}.

\bibitem[{United Nations}(2024)]{unsc2024}
{United Nations}.
\newblock {Security Council Fails to Adopt First-Ever Resolution on Arms Race
  in Outer Space, Due to Negative Vote by Russian Federation}.
\newblock SC/15678, 24 April, 2024.
\newblock URL \url{https://press.un.org/en/2024/sc15678.doc.htm}.

\bibitem[Hess(1963)]{hess1963}
Wilmot~N. Hess.
\newblock {The Artificial Radiation Belt}.
\newblock Technical Report NASA-TM-X-788, NASA Goddard Space Flight Center,
  1963.
\newblock URL
  \url{https://www.scribd.com/document/131532114/The-Artificial-Radiation-Belt-NASA-Technical-Memorandum-X-788-April-1963}.

\bibitem[Hoerlin(1976)]{hoerlin1976}
Herman Hoerlin.
\newblock {United States high-altitude test experiences: A review emphasizing
  the impact on the environment}.
\newblock Technical Report LA-6405, Los Alamos Scientific Laboratory, 1976.

\bibitem[Gombosi et~al.(2017)Gombosi, Baker, Balogh, Erickson, Huba, and
  Lanzerotti]{gombosi2017}
Tamas~I. Gombosi, Daniel~N. Baker, Andr{\'e} Balogh, Philip~J. Erickson,
  Joseph~D. Huba, and Louis~J. Lanzerotti.
\newblock {Anthropogenic Space Weather}.
\newblock \emph{Space Science Reviews}, 212\penalty0 (3--4):\penalty0
  985--1039, 2017.
\newblock \doi{10.1007/s11214-017-0357-5}.
\newblock URL \url{https://doi.org/10.1007/s11214-017-0357-5}.

\bibitem[Fang and Moric(2026)]{fang2026}
Zhengqing Fang and Igor Moric.
\newblock {Assessing the Impact of the Prompt Effect of a High-Altitude Nuclear
  Explosion on Non-Military {LEO} Satellites}.
\newblock \emph{Science \& Global Security}, 2026.

\bibitem[Cowee and Winske(2012)]{cowee2012}
Misa~M. Cowee and Dan Winske.
\newblock {Update on the Electron Source Model}.
\newblock Technical Report LA-UR-12-23076, Los Alamos National Laboratory,
  2012.
\newblock URL \url{https://www.osti.gov/servlets/purl/1046529}.

\bibitem[Cunningham and Cowee(2017)]{cunningham2017}
Gregory~S. Cunningham and Misa~M. Cowee.
\newblock {Incorporation of Drift Loss Cone Effects on the Trapping of an
  Artificial Radiation Belt into {LANL}'s Modeling Capability}.
\newblock \emph{Journal of Radiation Effects: Research and Engineering},
  35\penalty0 (1), 2017.
\newblock URL \url{https://www.osti.gov/servlets/purl/1485398}.

\bibitem[Thiele et~al.(2026)Thiele, Heiland, Boley, and Lawler]{thiele2026}
Sarah Thiele, Skye~R. Heiland, Aaron~C. Boley, and Samantha~M. Lawler.
\newblock {An orbital house of cards: Frequent satellite close conjunctions}.
\newblock \emph{Acta Astronautica}, 248:\penalty0 1289--1297, 2026.
\newblock \doi{10.1016/j.actaastro.2026.06.023}.
\newblock URL \url{https://doi.org/10.1016/j.actaastro.2026.06.023}.

\bibitem[Thiele and Boley(2022)]{thiele2022}
Sarah Thiele and Aaron~C. Boley.
\newblock {Investigating the Risks of Debris-Generating {ASAT} Tests in the
  Presence of Megaconstellations}.
\newblock \emph{The Journal of the Astronautical Sciences}, 69\penalty0
  (6):\penalty0 1797--1820, 2022.
\newblock \doi{10.1007/s40295-022-00356-6}.
\newblock URL \url{https://doi.org/10.1007/s40295-022-00356-6}.

\bibitem[de~Troullioud~de Lanversin et~al.(2021)de~Troullioud~de Lanversin,
  K{\"u}tt, and Glaser]{deTroullioud2021}
Julien de~Troullioud~de Lanversin, Moritz K{\"u}tt, and Alexander Glaser.
\newblock {ONIX: An open-source depletion code}.
\newblock \emph{Annals of Nuclear Energy}, 151:\penalty0 107903, 2021.

\bibitem[Beggan et~al.(2026)Beggan, Kloss, Amblard, Amezza, Amit, Aubert,
  Baerenzung, Bregnh{\o}j~Lauridsen, Brown, Califf, et~al.]{beggan2026}
C.~D. Beggan, Clemens Kloss, P~Amblard, T~Amezza, H~Amit, J~Aubert,
  J~Baerenzung, J~Bregnh{\o}j~Lauridsen, W~Brown, S~Califf, et~al.
\newblock {International geomagnetic reference field: the fourteenth
  generation}.
\newblock \emph{Earth, Planets and Space}, 78:\penalty0 127, 2026.
\newblock \doi{10.1186/s40623-025-02360-0}.

\bibitem[D'Arcy and Colgate(1965)]{dArcy1965}
Raymond~G. D'Arcy and Stirling~A. Colgate.
\newblock {Measurements at the southern magnetic conjugate region of the
  fission debris from the Starfish nuclear detonation}.
\newblock \emph{Journal of Geophysical Research}, 70\penalty0 (13):\penalty0
  3147--3159, 1965.

\bibitem[{U.S. Department of Defense}(2022)]{milprf38535}
{U.S. Department of Defense}.
\newblock {Integrated Circuits (Microcircuits) Manufacturing, General
  Specification for}.
\newblock Technical Report MIL-PRF-38535M, Defense Logistics Agency, November
  2022.
\newblock URL
  \url{https://quicksearch.dla.mil/qaDocDetails.aspx?ident_number=69354}.

\bibitem[Cladis et~al.(1971)Cladis, Davidson, and Newkirk]{cladis1971}
John~B. Cladis, Gerald~T. Davidson, and Lester~L. Newkirk, editors.
\newblock \emph{{The Trapped Radiation Handbook}}.
\newblock Defense Nuclear Agency, 1971.
\newblock URL \url{https://apps.dtic.mil/sti/tr/pdf/AD0738841.pdf}.

\bibitem[{Outer Space Institute}(2026)]{osi2026crash}
{Outer Space Institute}.
\newblock {CRASH Clock}, 2026.
\newblock URL \url{https://outerspaceinstitute.ca/crashclock/}.

\bibitem[Johnson et~al.(2001)Johnson, Krisko, Liou, and
  Anz-Meador]{johnson2001}
Nicholas~L. Johnson, Paula~H. Krisko, Jer-Chyi Liou, and Phillip~D. Anz-Meador.
\newblock {{NASA}'s new breakup model of {EVOLVE} 4.0}.
\newblock \emph{Advances in Space Research}, 28\penalty0 (9):\penalty0
  1377--1384, 2001.
\newblock \doi{10.1016/S0273-1177(01)00423-9}.

\bibitem[{NASA Marshall Space Flight Center}(2026)]{nasa2026}
{NASA Marshall Space Flight Center}.
\newblock {Solar Cycle Progression and Forecast}, 2026.
\newblock URL \url{https://www.nasa.gov/solar-cycle-progression-and-forecast/}.

\bibitem[{COSPAR and URSI}(2012)]{cira2012}
{COSPAR and URSI}.
\newblock {{COSPAR International Reference Atmosphere 2012}: Models of the
  Earth's Upper Atmosphere}.
\newblock Technical report, COSPAR and URSI, 2012.
\newblock URL
  \url{https://spacewx.com/wp-content/uploads/2021/03/chapters_1_3.pdf}.

\bibitem[Bostrom et~al.(1970)Bostrom, Beall, and Armstrong]{bostrom1970}
Carl~O. Bostrom, David~Strickland Beall, and James~C. Armstrong.
\newblock {Time history of the inner radiation zone, October 1963 to December
  1968}.
\newblock \emph{Journal of Geophysical Research}, 75\penalty0 (7):\penalty0
  1246--1256, 1970.
\newblock \doi{10.1029/JA075i007p01246}.

\bibitem[Lewis and Kessler(2025)]{lewis2025}
H.~G. Lewis and D.~J. Kessler.
\newblock Critical number of spacecraft in low {Earth} orbit: A new assessment
  of the stability of the orbital debris environment.
\newblock In \emph{Proceedings of the 9th European Conference on Space Debris},
  Bonn, Germany, 2025. ESA Space Debris Office.
\newblock URL
  \url{https://conference.sdo.esoc.esa.int/proceedings/sdc9/paper/305}.

\bibitem[Glasstone and Dolan(1977)]{glasstone1977}
Samuel Glasstone and Philip~J. Dolan.
\newblock \emph{{The Effects of Nuclear Weapons}}.
\newblock U.S. Department of Defense and U.S. Department of Energy, 3 edition,
  1977.
\newblock URL \url{https://fissilematerials.org/library/gla77.pdf}.

\bibitem[Petersen et~al.(2026)Petersen, Lockert, and Philippe]{petersen2026}
Zachary Petersen, Tristan Lockert, and S{\'e}bastien Philippe.
\newblock {Evaluating the Effects of a Nuclear Detonation in Low Earth Orbit}.
\newblock In \emph{Proceedings of the 67th Annual Meeting of the Institute of
  Nuclear Materials Management}, August 2026.

\bibitem[Murphy et~al.(2023)]{murphy2023}
Daniel~M. Murphy et~al.
\newblock Metals from spacecraft reentry in stratospheric aerosol particles.
\newblock \emph{Proceedings of the National Academy of Sciences}, 120\penalty0
  (43):\penalty0 e2313374120, 2023.
\newblock \doi{10.1073/pnas.2313374120}.
\newblock URL \url{https://pmc.ncbi.nlm.nih.gov/articles/PMC10614211/}.

\bibitem[Wang et~al.(2025)]{wang2025a}
Yanbin Wang et~al.
\newblock {Improvement on Analysis Method for X-Ray Arrival Spectrum From
  High-Altitude Nuclear Detonation}.
\newblock \emph{{IEEE} Transactions on Nuclear Science}, 72\penalty0
  (5):\penalty0 1706--1717, 2025.
\newblock \doi{10.1109/TNS.2025.3554347}.
\newblock URL \url{https://doi.org/10.1109/TNS.2025.3554347}.

\bibitem[Zhu et~al.(2024)Zhu, Zhang, Li, Li, Wang, and Zhang]{zhu2024}
Ding-Han Zhu, Xiong Zhang, Xiao-Qiang Li, Peng Li, Yan-Bin Wang, and Shuang
  Zhang.
\newblock {Atmospheric transmission algorithm for pulsed X-rays from
  high-altitude nuclear detonations based on scattering correction}.
\newblock \emph{Nuclear Science and Techniques}, 35\penalty0 (3):\penalty0 52,
  2024.
\newblock \doi{10.1007/s41365-024-01411-7}.
\newblock URL \url{https://doi.org/10.1007/s41365-024-01411-7}.

\bibitem[{National Oceanic and Atmospheric Administration}
  et~al.(1976){National Oceanic and Atmospheric Administration}, {National
  Aeronautics and Space Administration}, and {United States Air
  Force}]{noaa1976}
{National Oceanic and Atmospheric Administration}, {National Aeronautics and
  Space Administration}, and {United States Air Force}.
\newblock {U.S. Standard Atmosphere, 1976}.
\newblock Technical Report NOAA-S/T-76-1562, U.S. Government Printing Office,
  1976.
\newblock URL \url{https://ntrs.nasa.gov/citations/19770009539}.

\bibitem[Brown et~al.(2018)Brown, Chadwick, Capote, Kahler, Trkov, Herman,
  Sonzogni, Danon, Carlson, Dunn, et~al.]{brown2018}
David~A Brown, Mark~Benjamin Chadwick, Roberto Capote, AC~Kahler, A~Trkov,
  MW~Herman, AA~Sonzogni, Y~Danon, AD~Carlson, M~Dunn, et~al.
\newblock {{ENDF/B-VIII.0}: The 8th Major Release of the Nuclear Reaction Data
  Library with {CIELO}-project Cross Sections, New Standards and Thermal
  Scattering Data}.
\newblock \emph{Nuclear Data Sheets}, 148:\penalty0 1--142, 2018.
\newblock \doi{10.1016/j.nds.2018.02.001}.

\bibitem[Carter et~al.(1959)Carter, Reines, Wagner, and Wyman]{carter1959}
Robert~E. Carter, Frederick Reines, John~J. Wagner, and Marvin~Eugene Wyman.
\newblock {Free Antineutrino Absorption Cross Section. {II}. Expected Cross
  Section from Measurements of Fission Fragment Electron Spectrum}.
\newblock \emph{Physical Review}, 113:\penalty0 280--286, 1959.
\newblock \doi{10.1103/PhysRev.113.280}.

\bibitem[Brautigam and Albert(2000)]{brautigam2000}
Donald~H. Brautigam and Jay~M. Albert.
\newblock {Radial diffusion analysis of outer radiation belt electrons during
  the October 9, 1990, magnetic storm}.
\newblock \emph{Journal of Geophysical Research: Space Physics}, 105\penalty0
  (A1):\penalty0 291--309, 2000.
\newblock \doi{10.1029/1999JA900344}.
\newblock URL \url{https://doi.org/10.1029/1999JA900344}.

\bibitem[Reeves et~al.(2012)]{reeves2012}
Geoffrey~D. Reeves et~al.
\newblock {Dynamic Radiation Environment Assimilation Model: {DREAM}}.
\newblock \emph{Space Weather}, 10\penalty0 (3):\penalty0 S03006, 2012.
\newblock \doi{10.1029/2011SW000729}.
\newblock URL \url{https://doi.org/10.1029/2011SW000729}.

\bibitem[Lee et~al.(2024)]{lee2024}
Sang-Yun Lee et~al.
\newblock {Simulating Long-Term Dynamics of Radiation Belt Electrons Using
  {DREAM}3D Model}.
\newblock \emph{Journal of Geophysical Research: Space Physics}, 129\penalty0
  (2):\penalty0 e2023JA032286, 2024.
\newblock \doi{10.1029/2023JA032286}.
\newblock URL
  \url{https://gfzpublic.gfz.de/pubman/item/item_5025471_5/component/file_5025510/5025471.pdf}.

\bibitem[Berger et~al.(2017)Berger, Coursey, Zucker, and Chang]{nistestar}
Martin~J. Berger, Johnathan~S. Coursey, M.~A. Zucker, and J.~Chang.
\newblock {{ESTAR, PSTAR, and ASTAR}: Computer Programs for Calculating
  Stopping-Power and Range Tables}.
\newblock NIST Standard Reference Database 124, 2017.
\newblock URL \url{https://physics.nist.gov/Star}.

\bibitem[{Space Exploration Holdings, LLC}(2022)]{spacex2022geometry}
{Space Exploration Holdings, LLC}.
\newblock Supplemental information on the {Starlink Gen2} system: {FCC} filing
  of 4 october 2022, exhibit b.
\newblock IBFS File Nos. SAT-LOA-20200526-00055 and SAT-AMD-20210818-00105,
  October 2022.
\newblock URL \url{https://planet4589.org/astro/starsim/papers/StarGen2.pdf}.

\bibitem[McDowell(2026)]{gcat2026}
Jonathan~C. McDowell.
\newblock {{GCAT}: General Catalog of Artificial Space Objects}, 2026.
\newblock URL \url{https://planet4589.org/space/gcat/}.

\bibitem[Garc{\'i}a-Pelayo et~al.(2018)Garc{\'i}a-Pelayo, Gonzalo, and
  Bombardelli]{garciapelayo2018}
Ricardo Garc{\'i}a-Pelayo, Juan~Luis Gonzalo, and Claudio Bombardelli.
\newblock Rate and collision probability of tethers and sails against debris or
  spacecraft.
\newblock In \emph{7th International Conference on Astrodynamics Tools and
  Techniques (ICATT)}, DLR Oberpfaffenhofen, Germany, 2018.
\newblock URL
  \url{https://indico.esa.int/event/224/papers/4115/files/199-7ICATTPaperGarcia-Pelayo.pdf}.

\bibitem[Sutton et~al.(2025)Sutton, Fitzpatrick, Wang, and Fang]{sutton2025}
Eric~K. Sutton, David Fitzpatrick, Nai-Yu Wang, and Tzu-Wei Fang.
\newblock {Satellite Constellation Data to Drive Thermospheric Density
  Forecasting Capabilities}.
\newblock Space Weather Workshop poster, 2025.
\newblock URL
  \url{https://cpaess.ucar.edu/sites/default/files/poster-2025-sww/Sutton-Eric.pdf}.

\bibitem[{SpaceX}(2024)]{spacex2024}
{SpaceX}.
\newblock {Starlink Progress 2024}.
\newblock Technical report, SpaceX, 2024.
\newblock URL
  \url{https://www.starlink.com/public-files/starlinkProgressReport_2024.pdf}.

\bibitem[{ESA Space Debris Mitigation Working Group}(2025)]{esa2025}
{ESA Space Debris Mitigation Working Group}.
\newblock {ESA Space Debris Mitigation Compliance Verification Guidelines}.
\newblock Technical Report ESSB-HB-U-002, European Space Agency, 2025.
\newblock URL
  \url{https://sdup.esoc.esa.int/documents/download/ESA_Space_Debris_Mitigation_Compliance_Verification_Guidelines.pdf}.

\bibitem[Fitzpatrick(2012)]{fitzpatrick}
Richard Fitzpatrick.
\newblock \emph{An Introduction to Celestial Mechanics}.
\newblock Cambridge University Press, 2012.

\end{thebibliography}


\begin{thebibliography}{24}
\providecommand{\natexlab}[1]{#1}
\providecommand{\url}[1]{\texttt{#1}}
\expandafter\ifx\csname urlstyle\endcsname\relax
  \providecommand{\doi}[1]{doi: #1}\else
  \providecommand{\doi}{doi: \begingroup \urlstyle{rm}\Url}\fi

\bibitem[Kelso(2026)]{supp-kelso2026}
Thomas~Sean Kelso.
\newblock {{CelesTrak}: Current {NORAD} Two-Line Element Sets}, 2026.
\newblock URL \url{https://celestrak.org/NORAD/elements/}.

\bibitem[Sutton et~al.(2025)Sutton, Fitzpatrick, Wang, and
  Fang]{supp-sutton2025}
Eric~K. Sutton, David Fitzpatrick, Nai-Yu Wang, and Tzu-Wei Fang.
\newblock {Satellite Constellation Data to Drive Thermospheric Density
  Forecasting Capabilities}.
\newblock Space Weather Workshop poster, 2025.
\newblock URL
  \url{https://cpaess.ucar.edu/sites/default/files/poster-2025-sww/Sutton-Eric.pdf}.

\bibitem[{SpaceX}(2024)]{supp-spacex2024}
{SpaceX}.
\newblock {Starlink Progress 2024}.
\newblock Technical report, SpaceX, 2024.
\newblock URL
  \url{https://www.starlink.com/public-files/starlinkProgressReport_2024.pdf}.

\bibitem[McDowell(2026)]{supp-gcat2026}
Jonathan~C. McDowell.
\newblock {{GCAT}: General Catalog of Artificial Space Objects}, 2026.
\newblock URL \url{https://planet4589.org/space/gcat/}.

\bibitem[{National Aeronautics and Space Administration}(2026)]{supp-nasaISS}
{National Aeronautics and Space Administration}.
\newblock {International Space Station}: Facts and figures, 2026.
\newblock URL
  \url{https://www.nasa.gov/international-space-station/space-station-facts-and-figures/}.

\bibitem[de~Troullioud~de Lanversin et~al.(2021)de~Troullioud~de Lanversin,
  K{\"u}tt, and Glaser]{supp-deTroullioud2021}
Julien de~Troullioud~de Lanversin, Moritz K{\"u}tt, and Alexander Glaser.
\newblock {ONIX: An open-source depletion code}.
\newblock \emph{Annals of Nuclear Energy}, 151:\penalty0 107903, 2021.

\bibitem[{NASA Marshall Space Flight Center}(2026)]{supp-nasa2026}
{NASA Marshall Space Flight Center}.
\newblock {Solar Cycle Progression and Forecast}, 2026.
\newblock URL \url{https://www.nasa.gov/solar-cycle-progression-and-forecast/}.

\bibitem[{COSPAR and URSI}(2012)]{supp-cira2012}
{COSPAR and URSI}.
\newblock {{COSPAR International Reference Atmosphere 2012}: Models of the
  Earth's Upper Atmosphere}.
\newblock Technical report, COSPAR and URSI, 2012.
\newblock URL
  \url{https://spacewx.com/wp-content/uploads/2021/03/chapters_1_3.pdf}.

\bibitem[Beggan et~al.(2026)Beggan, Kloss, Amblard, Amezza, Amit, Aubert,
  Baerenzung, Bregnh{\o}j~Lauridsen, Brown, Califf, et~al.]{supp-beggan2026}
C.~D. Beggan, Clemens Kloss, P~Amblard, T~Amezza, H~Amit, J~Aubert,
  J~Baerenzung, J~Bregnh{\o}j~Lauridsen, W~Brown, S~Califf, et~al.
\newblock {International geomagnetic reference field: the fourteenth
  generation}.
\newblock \emph{Earth, Planets and Space}, 78:\penalty0 127, 2026.
\newblock \doi{10.1186/s40623-025-02360-0}.

\bibitem[Brown et~al.(2018)Brown, Chadwick, Capote, Kahler, Trkov, Herman,
  Sonzogni, Danon, Carlson, Dunn, et~al.]{supp-brown2018}
David~A Brown, Mark~Benjamin Chadwick, Roberto Capote, AC~Kahler, A~Trkov,
  MW~Herman, AA~Sonzogni, Y~Danon, AD~Carlson, M~Dunn, et~al.
\newblock {{ENDF/B-VIII.0}: The 8th Major Release of the Nuclear Reaction Data
  Library with {CIELO}-project Cross Sections, New Standards and Thermal
  Scattering Data}.
\newblock \emph{Nuclear Data Sheets}, 148:\penalty0 1--142, 2018.
\newblock \doi{10.1016/j.nds.2018.02.001}.

\bibitem[D'Arcy and Colgate(1965)]{supp-dArcy1965}
Raymond~G. D'Arcy and Stirling~A. Colgate.
\newblock {Measurements at the southern magnetic conjugate region of the
  fission debris from the Starfish nuclear detonation}.
\newblock \emph{Journal of Geophysical Research}, 70\penalty0 (13):\penalty0
  3147--3159, 1965.

\bibitem[Carter et~al.(1959)Carter, Reines, Wagner, and Wyman]{supp-carter1959}
Robert~E. Carter, Frederick Reines, John~J. Wagner, and Marvin~Eugene Wyman.
\newblock {Free Antineutrino Absorption Cross Section. {II}. Expected Cross
  Section from Measurements of Fission Fragment Electron Spectrum}.
\newblock \emph{Physical Review}, 113:\penalty0 280--286, 1959.
\newblock \doi{10.1103/PhysRev.113.280}.

\bibitem[Cowee and Winske(2012)]{supp-cowee2012}
Misa~M. Cowee and Dan Winske.
\newblock {Update on the Electron Source Model}.
\newblock Technical Report LA-UR-12-23076, Los Alamos National Laboratory,
  2012.
\newblock URL \url{https://www.osti.gov/servlets/purl/1046529}.

\bibitem[Brautigam and Albert(2000)]{supp-brautigam2000}
Donald~H. Brautigam and Jay~M. Albert.
\newblock {Radial diffusion analysis of outer radiation belt electrons during
  the October 9, 1990, magnetic storm}.
\newblock \emph{Journal of Geophysical Research: Space Physics}, 105\penalty0
  (A1):\penalty0 291--309, 2000.
\newblock \doi{10.1029/1999JA900344}.
\newblock URL \url{https://doi.org/10.1029/1999JA900344}.

\bibitem[Reeves et~al.(2012)]{supp-reeves2012}
Geoffrey~D. Reeves et~al.
\newblock {Dynamic Radiation Environment Assimilation Model: {DREAM}}.
\newblock \emph{Space Weather}, 10\penalty0 (3):\penalty0 S03006, 2012.
\newblock \doi{10.1029/2011SW000729}.
\newblock URL \url{https://doi.org/10.1029/2011SW000729}.

\bibitem[Lee et~al.(2024)]{supp-lee2024}
Sang-Yun Lee et~al.
\newblock {Simulating Long-Term Dynamics of Radiation Belt Electrons Using
  {DREAM}3D Model}.
\newblock \emph{Journal of Geophysical Research: Space Physics}, 129\penalty0
  (2):\penalty0 e2023JA032286, 2024.
\newblock \doi{10.1029/2023JA032286}.
\newblock URL
  \url{https://gfzpublic.gfz.de/pubman/item/item_5025471_5/component/file_5025510/5025471.pdf}.

\bibitem[{Space Exploration Holdings, LLC}(2022)]{supp-spacex2022geometry}
{Space Exploration Holdings, LLC}.
\newblock Supplemental information on the {Starlink Gen2} system: {FCC} filing
  of 4 october 2022, exhibit b.
\newblock IBFS File Nos. SAT-LOA-20200526-00055 and SAT-AMD-20210818-00105,
  October 2022.
\newblock URL \url{https://planet4589.org/astro/starsim/papers/StarGen2.pdf}.

\bibitem[Garc{\'i}a-Pelayo et~al.(2018)Garc{\'i}a-Pelayo, Gonzalo, and
  Bombardelli]{supp-garciapelayo2018}
Ricardo Garc{\'i}a-Pelayo, Juan~Luis Gonzalo, and Claudio Bombardelli.
\newblock Rate and collision probability of tethers and sails against debris or
  spacecraft.
\newblock In \emph{7th International Conference on Astrodynamics Tools and
  Techniques (ICATT)}, DLR Oberpfaffenhofen, Germany, 2018.
\newblock URL
  \url{https://indico.esa.int/event/224/papers/4115/files/199-7ICATTPaperGarcia-Pelayo.pdf}.

\bibitem[Thiele et~al.(2026)Thiele, Heiland, Boley, and
  Lawler]{supp-thiele2026}
Sarah Thiele, Skye~R. Heiland, Aaron~C. Boley, and Samantha~M. Lawler.
\newblock {An orbital house of cards: Frequent satellite close conjunctions}.
\newblock \emph{Acta Astronautica}, 248:\penalty0 1289--1297, 2026.
\newblock \doi{10.1016/j.actaastro.2026.06.023}.
\newblock URL \url{https://doi.org/10.1016/j.actaastro.2026.06.023}.

\bibitem[Thiele and Boley(2022)]{supp-thiele2022}
Sarah Thiele and Aaron~C. Boley.
\newblock {Investigating the Risks of Debris-Generating {ASAT} Tests in the
  Presence of Megaconstellations}.
\newblock \emph{The Journal of the Astronautical Sciences}, 69\penalty0
  (6):\penalty0 1797--1820, 2022.
\newblock \doi{10.1007/s40295-022-00356-6}.
\newblock URL \url{https://doi.org/10.1007/s40295-022-00356-6}.

\bibitem[Johnson et~al.(2001)Johnson, Krisko, Liou, and
  Anz-Meador]{supp-johnson2001}
Nicholas~L. Johnson, Paula~H. Krisko, Jer-Chyi Liou, and Phillip~D. Anz-Meador.
\newblock {{NASA}'s new breakup model of {EVOLVE} 4.0}.
\newblock \emph{Advances in Space Research}, 28\penalty0 (9):\penalty0
  1377--1384, 2001.
\newblock \doi{10.1016/S0273-1177(01)00423-9}.

\bibitem[Fitzpatrick et~al.(2026)Fitzpatrick, Sutton, Pilinski, and
  Palo]{supp-fitzpatrick2026}
David~J. Fitzpatrick, Eric~K. Sutton, Marcin~D. Pilinski, and Scott~E. Palo.
\newblock {Harnessing Satellite Constellations as Signals of Opportunity for an
  Enhanced Specification of the Satellite Drag Environment in the Critical
  {LEO} Regime}.
\newblock \emph{Space Weather}, 24\penalty0 (3):\penalty0 e2025SW004611, 2026.
\newblock \doi{10.1029/2025SW004611}.

\bibitem[{ESA Space Debris Mitigation Working Group}(2025)]{supp-esa2025}
{ESA Space Debris Mitigation Working Group}.
\newblock {ESA Space Debris Mitigation Compliance Verification Guidelines}.
\newblock Technical Report ESSB-HB-U-002, European Space Agency, 2025.
\newblock URL
  \url{https://sdup.esoc.esa.int/documents/download/ESA_Space_Debris_Mitigation_Compliance_Verification_Guidelines.pdf}.

\bibitem[Fitzpatrick(2012)]{supp-fitzpatrick}
Richard Fitzpatrick.
\newblock \emph{An Introduction to Celestial Mechanics}.
\newblock Cambridge University Press, 2012.

\end{thebibliography}

\clearpage

\begingroup
\normalfont\normalsize\singlespacing

\setlength{\belowcaptionskip}{8pt}
\captionsetup{font=small,skip=12pt}
\captionsetup[figure]{position=bottom}
\captionsetup[table]{position=top}
\setlength{\tabcolsep}{7pt}
\renewcommand{\arraystretch}{1.2}

\begin{center}
{\large\bfseries Supplementary Information}\par
\vspace{0.8em}
{\Large\bfseries Effects of a Nuclear Detonation in Low Earth Orbit}\par
\vspace{1.2em}
Zachary Petersen$^{1}$, Hong Qin$^{1}$, Sarah Thiele$^{2}$,\par
and S\'{e}bastien Philippe$^{1,*}$\par
\vspace{0.8em}
{\small
$^{1}$Department of Nuclear Engineering and Engineering Physics,\par
University of Wisconsin--Madison, Madison, WI, USA\par
\vspace{0.4em}
$^{2}$Department of Astrophysical Sciences,\par
Princeton University, Princeton, NJ, USA\par
\vspace{0.6em}
$^{*}$Correspondence: \href{mailto:sphilippe@wisc.edu}{sphilippe@wisc.edu}\par
}
\end{center}
\vspace{1.4em}
\noindent\textbf{Contents}\par
\medskip
\noindent This supplement provides additional figures, model inputs, sensitivity results and detailed methods supporting the radiation, collision and orbital-decay calculations.
\begin{itemize}
\item \textbf{Supplementary Figures S1--S2:} the altitude distribution of the radiation-modeled population and the collision hazard for an inserted satellite relative to the no-detonation model.
\item \textbf{Supplementary Tables S1--S4:} adopted spacecraft masses and drag properties; radiation-driven fleet losses across electron-source, tolerance and yield assumptions; and ten-year collision and debris sensitivity results.
\item \textbf{Supplementary Methods:} catalogue selection, geomagnetic field representation, electron production, debris expansion, artificial-belt evolution, collision geometry and accounting, spacecraft drag properties, fragmentation and mass conservation, fragment transport, atmospheric forcing, numerical checks and the inserted-satellite calculation.
\end{itemize}

\clearpage

\setcounter{figure}{0}
\renewcommand{\thefigure}{S\arabic{figure}}

\begin{figure}[!htbp]
  \centering
  \includegraphics[width=0.82\textwidth]{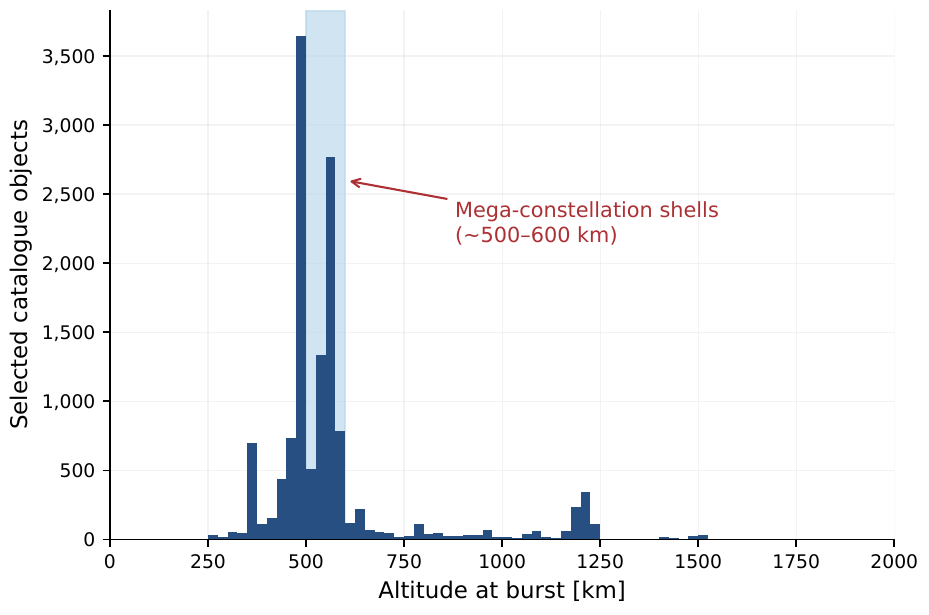}

\caption{\textbf{Altitude distribution of the objects used for the radiation calculation.} Counts of the 13,306 selected objects at 08:00 UTC on 14 January 2026, propagated from the retained CelesTrak elements~\protect\citesupp{supp-kelso2026}. Bars use 25-km bins; the blue band highlights 500--600~km. Altitude is evaluated at the burst time.}
\label{fig:leo_population}

\end{figure}
\clearpage

\begin{figure}[!htbp]
  \centering
  \includegraphics[width=0.97\textwidth]{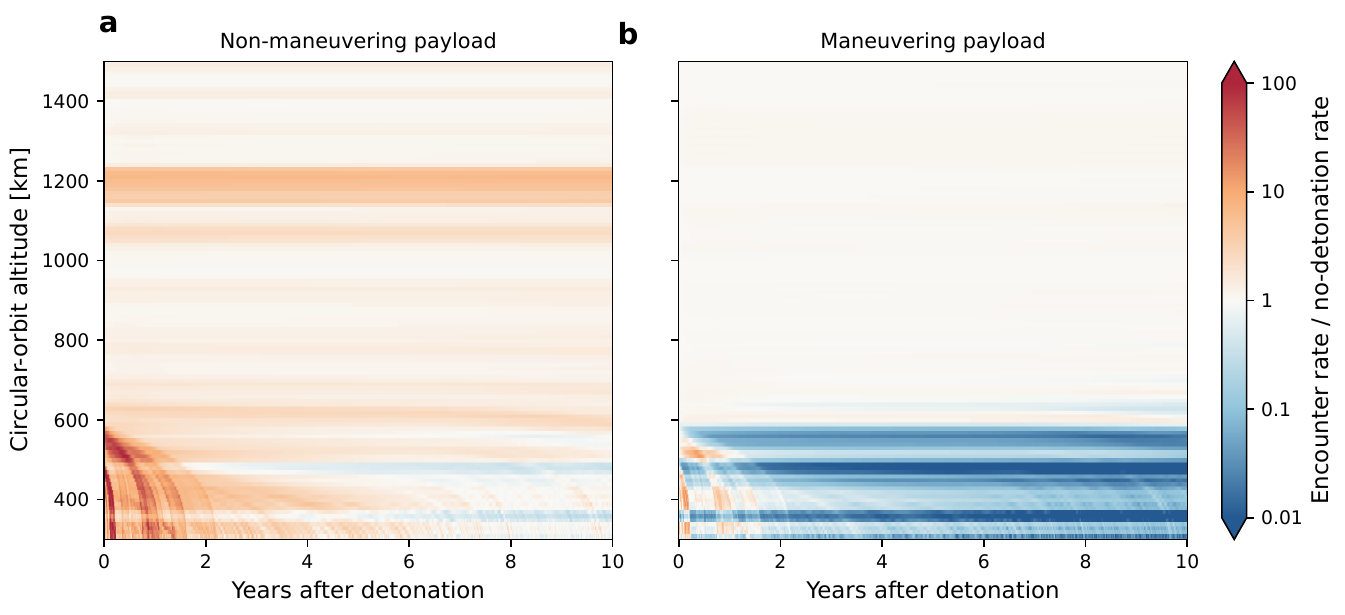}

\caption{\textbf{Altitude and time determine the relative collision hazard for an inserted satellite.} Instantaneous rate for a payload in a circular orbit divided by its rate in the matched no-detonation model. Both calculations use the same $9\times2.8\times0.2$-m geometric proxy. \textbf{(a)} Non-maneuvering payload, with avoidance credit only against operational partners. \textbf{(b)} Maneuvering payload, with an assumed 99\% avoidance benefit against every modeled partner. Blue denotes a ratio below one and red above one; white denotes equal rates. Triangular color-bar extensions mark values outside 0.01--100. Gray cells, if present, have no positive control rate. Both models use the January inventory, central solar forcing and identical test geometry. Radiation failure of the test payload and debris below 10~cm are excluded. Rates below the control reflect fewer collision targets and do not establish that operations could safely resume.}
\label{fig:orbit_collision_rate_relative_to_background}

\end{figure}
\clearpage

\setcounter{table}{0}
\renewcommand{\thetable}{S\arabic{table}}

\clearpage
\begin{table}[!htbp]
\centering\small

\caption{\textbf{Adopted masses and drag properties.} Masses are in kg and area-to-mass ratios in m$^2$\,kg$^{-1}$. Starlink values are generation-specific proxies~\protect\citesupp{supp-sutton2025,supp-spacex2024,supp-gcat2026}. Other objects use available GCAT dry masses, with the stated fallback when unavailable. Docked complexes are counted once: the ISS uses a 420-tonne nominal proxy~\protect\citesupp{supp-nasaISS}; Tiangong uses summed component dry masses. Their projected geometric areas also serve as drag-area proxies. Collision geometry is assigned separately (Methods).}
\label{tab:intact_properties}
\setlength{\tabcolsep}{4pt}
\begin{tabular}{lrrrr}
\hline
Class & Mass & Central $A/m$ & Lower & Upper \\ \hline
Starlink v1.0 & 260 & 0.046 & 0.035 & 0.060 \\
Starlink v1.5 & 306 & 0.044 & 0.035 & 0.060 \\
Starlink V2 Mini & 800 & 0.070 & 0.050 & 0.100 \\
Starlink optimized & 575 & 0.100 & 0.0696 & 0.1391 \\
Starlink direct-to-cell & 960 & 0.0604 & 0.0417 & 0.0833 \\
Other payloads & GCAT / 260 & 0.010 & --- & --- \\
Rocket bodies & GCAT / 1000 & 0.005 & --- & --- \\
\hline
\end{tabular}

\end{table}

\clearpage
\begin{table}[!htbp]
  \centering
  \small
  \caption{\textbf{Radiation-driven fleet loss across electron-source and tolerance assumptions.}
  Percentage of the 13,306-object fleet disabled after a 1-Mt detonation
  at 500~km over $20^{\circ}$S, $35^{\circ}$W, behind 2~mm of aluminium.
  The source is emitted electrons per fission before the calculated
  trapping efficiency is applied. Source counts are derived from
  ONIX beta-activity histories~\protect\citesupp{supp-deTroullioud2021}.
  Losses include prompt fluence of at least 40~J\,m$^{-2}$ or
  prompt-plus-belt dose reaching the assigned threshold. The final
  column gives the first three-hour sample at which 90\% of the fleet
  has crossed the failure criterion. Prompt X-rays alone disable 2.9\% at $t=0$.
  }
  \label{tab:radiation_loss_cases}
  \begin{tabular}{rrrrrr}
    \hline
    Tolerance & Electrons & Day 7 & Day 15 & Day 30 & $t_{90\%}$ \\
    {}[krad(Si)] & per fission & [\%] & [\%] & [\%] & [days] \\
    \hline
40 & 0.404 & 67.2 & 96.8 & 99.2 & 10.38 \\
40 & 1.289 & 97.8 & 99.4 & 99.8 & 4.12 \\
40 & 5.207 & 99.2 & 99.8 & 99.9 & 2.12 \\

120 & 0.404 & 11.5 & 37.4 & 90.0 & 30.00 \\
120 & 1.289 & 71.8 & 97.2 & 99.3 & 9.88 \\
120 & 5.207 & 98.2 & 99.4 & 99.8 & 4.00 \\
    \hline
  \end{tabular}
\end{table}
\clearpage
\begin{table}[!htbp]
  \centering
  \small
  \caption{\textbf{Radiation-driven fleet loss for 1-Mt, 100-kt and 10-kt detonations.}
  Percentage of the 13,306-object fleet disabled after a detonation at 500~km over
  $20^{\circ}$S, $35^{\circ}$W, with a 50\% fission fraction and 2~mm of aluminium shielding.
  Results use the central source of 1.29 emitted electrons per fission and two
  radiation-tolerance thresholds. Losses include prompt fluence
  of at least 40~J\,m$^{-2}$ or prompt-plus-belt dose reaching the threshold.
  Prompt X-rays alone disable 2.9\%, 0.3\% and $<$0.1\% of the fleet for 1~Mt, 100~kt
  and 10~kt, respectively. Smaller bursts populate narrower shell ranges:
  $L=1.31$--1.51, 1.33--1.41 and 1.34--1.38, respectively.
  The $t_{50\%}$ and $t_{90\%}$ values give the first three-hour samples at which
  50\% and 90\% of the fleet cross the failure criterion.}
  \label{tab:yield_loss}
  \setlength{\tabcolsep}{4pt}
  \begin{tabular}{lrrrrrrrrr}
    \hline
    Yield & Tolerance & Day 1 & Day 7 & Day 30 & Day 60 & Day 180 & Day 365 & $t_{50\%}$ & $t_{90\%}$ \\
    & [krad(Si)] & [\%] & [\%] & [\%] & [\%] & [\%] & [\%] & [days] & [days] \\
    \hline
    1 Mt & 40  & 25.2 & 97.8 & 99.8 & 99.9 & 99.9 & 99.9 & 1.9 & 4.1 \\
    1 Mt & 120 & 6.4  & 71.8 & 99.3 & 99.7 & 99.8 & 99.8 & 5.2 & 9.9 \\
    \hline
    100 kt & 40  & 0.5 & 11.2 & 86.8 & 98.6 & 99.3 & 99.4 & 17.4 & 33.2 \\
    100 kt & 120 & 0.3 & 1.8  & 14.2 & 64.7 & 96.5 & 96.7 & 51.6 & 92.9 \\
    \hline
    10 kt & 40  & 0.0 & 0.1 & 4.5 & 9.3 & 53.0 & 92.2 & 174.8 & 320.2 \\
    10 kt & 120 & 0.0 & 0.0 & 0.1 & 1.2 & 9.1 & 18.6 & $>365$ & $>365$ \\
    \hline
  \end{tabular}
\end{table}
\clearpage
\begin{table}[!htbp]
\centering\small

\caption{\textbf{Sensitivity of ten-year collision and debris outcomes.} The central calculation uses object-dependent geometric cross-sections, the central NASA/MSFC solar-flux path and the Starlink properties in Table~\ref{tab:intact_properties}. All cases use the same January inventory. Detonation cases use the same 40~krad(Si) radiation histories; the no-detonation control excludes radiation failures. Each sensitivity changes the stated assumption. Solar paths vary atmospheric forcing~\protect\citesupp{supp-nasa2026,supp-cira2012}; Starlink cases vary area-to-mass ratios and, in the final drag case, Starlink $C_D$ only. Fixed-area prescriptions replace geometric cross-sections with the encounter areas described in Methods. Fixed-density cases hold one CIRA profile throughout the decade. Collision totals include all pair types and catastrophic and cratering outcomes. Fragment counts refer to new fragments at least 10~cm in characteristic length; both population columns are weighted by orbital residence below 2000~km at year ten. These are sensitivity cases, not confidence intervals.}
\label{tab:collision_sensitivity}
\setlength{\tabcolsep}{5pt}
\begin{tabular}{lrrr}
\hline
Case & Expected collisions & New fragments & Disabled payloads \\
\hline
Central geometric case & 26.1 & 3,943 & 2,110 \\
No detonation, geometric & 12.4 & 2,650 & 0 \\
\hline
Solar 5th-index path & 28.2 & 5,518 & 2,402 \\
Solar 95th-index path & 24.5 & 2,884 & 1,741 \\
\hline
Lower Starlink $A/m$ & 34.8 & 4,170 & 2,110 \\
Upper Starlink $A/m$ & 20.0 & 3,781 & 2,110 \\
Upper $A/m$, $C_D=2.7$ & 17.3 & 3,709 & 2,110 \\
\hline
4.8-2.4-10 fixed areas & 16.4 & 3,150 & 2,107 \\
10-5-10 fixed areas & 88.4 & 14,136 & 2,084 \\
\hline
Fixed low density & 476.7 & 97,035 & 6,794 \\
Fixed moderate density & 23.7 & 3,403 & 1,881 \\
Fixed high density & 9.5 & 2,014 & 1,431 \\
\hline
\end{tabular}

\end{table}
\clearpage
\section*{Supplementary Methods}
\setcounter{equation}{0}
\renewcommand{\theequation}{S\arabic{equation}}
\renewcommand{\theHequation}{S\arabic{equation}}

The following sections provide the population selection rules and detailed radiation and collision-geometry calculations summarized in Methods.

\subsection*{Orbital population}

The radiation calculation uses General Perturbations element sets from 14 January 2026~\protect\citesupp{supp-kelso2026}. All objects are propagated with SGP4 to the burst epoch. Objects with successful propagation and instantaneous spherical altitudes greater than zero and no greater than 2000~km are retained, giving 13,306 objects. This selection is based on position at the burst time rather than requiring the entire orbit to remain within LEO.

The collision calculation combines the radiation catalogue with the full Space-Track catalogue for 14 January 2026. NORAD identifiers are deduplicated, with the radiation entries taking precedence. Selection requires a valid bound orbit, elements no more than 100 days old and positive orbital residence below 2000~km. The merged inventory contains 26,245 entries: 13,300 dose-tracked entries treated as payloads, 20 additional dose-unmodeled active-list payloads, 1754 other payloads, 1555 rocket bodies and 9616 debris objects. Of the 13,306 objects in the radiation calculation, six are classified as rocket bodies in the collision inventory. Radiation-loss fractions use the full 13,306-object population. Objects with unresolved types are listed in the catalogue audit and excluded from the collision inventory.

Nine ISS entries and six Tiangong entries are represented as two physical collision identities, leaving 16,616 intact collision identities. Components remain in a mapping ledger. Objects without a payload dose history receive no avoidance credit. No launches or scheduled disposals occur in either the detonation or control run. Shell counts are weighted by Keplerian orbital residence, so LEO-equivalent populations differ from identity counts. The radiation selection uses instantaneous burst altitude, whereas the additional collision inventory uses orbital residence; these selections serve different purposes.

\subsection*{Geomagnetic field}

The radiation calculation uses IGRF-14, evaluated for 14 January 2026~\protect\citesupp{supp-beggan2026}. The field is obtained from the degree-13 spherical-harmonic potential,
\begin{equation}
\begin{aligned}
\mathbf B &= -\nabla V,\\
V(r,\theta,\phi) &= a\sum_{n=1}^{13}\left(\frac{a}{r}\right)^{n+1}
\sum_{m=0}^{n}\left(g_n^m\cos m\phi+h_n^m\sin m\phi\right)P_n^m(\cos\theta),
\end{aligned}
\label{eq:igrf}
\end{equation}
where $a=6371.2$~km, $g_n^m$ and $h_n^m$ are the IGRF coefficients, and $P_n^m$ are Schmidt semi-normalized associated Legendre functions. Field lines are traced in both directions using fourth-order Runge--Kutta integration.

Shells are labelled by their minimum field strength,
\begin{equation}
L=\left(\frac{B_0}{B_{\min}}\right)^{1/3},
\qquad B_0=31,200~\mathrm{nT}.
\label{eq:lshell}
\end{equation}
This is a dipole-equivalent shell label rather than an exact invariant coordinate in the non-dipolar field. Shell properties are tabulated over 72 longitudes separated by $5^{\circ}$. The atmospheric loss boundary for trapping is 120~km. In the adopted 2026 field, shells below approximately $L=1.15$ do not support trapping under this prescription because their drift paths encounter the atmospheric boundary over the South Atlantic Anomaly. The satellite shell lookup grid covers 100--2100~km. Centered-dipole field lines are used only for the geometric projection in Fig.~\ref{fig:belt_geometry}; injection shells and trapping are calculated with IGRF-14.

\subsection*{Fission-electron source term}

The number of fissions follows from the yield $Y_{\mathrm{kt}}$, fission fraction $f_{\mathrm{fis}}=0.5$, and an assumed 180~MeV released per fission:
\begin{equation}
N_{\mathrm{fis}}=Y_{\mathrm{kt}}f_{\mathrm{fis}}
\frac{4.184\times10^{12}~\mathrm{J\,kt^{-1}}}
{180~\mathrm{MeV}\times1.602\times10^{-13}~\mathrm{J\,MeV^{-1}}}.
\label{eq:nfis}
\end{equation}
The reference detonation therefore produces approximately $7.25\times10^{25}$ fissions.

Electron production is obtained from time-dependent U-235 fission-product beta-activity tables calculated with the open-source depletion code ONIX~\protect\citesupp{supp-deTroullioud2021}, using ENDF/B-VIII.0 evaluated nuclear data~\protect\citesupp{supp-brown2018}. The activity $A_{\beta}(t)$ is interpolated linearly below 1~s and logarithmically in both time and activity above 1~s. Beyond the tabulated interval of 0--30 days, activity is continued as $A_{\beta}\propto t^{-1.118}$, fitted to the tabulated activity at times greater than one hour. The cumulative emission per fission is
\begin{equation}
n_{\beta}(t)=\int_0^t A_{\beta}(t')\,dt'.
\label{eq:betacumulative}
\end{equation}

Three source cases represent different durations for which radioactive debris remains available to supply electrons: 0.40 electrons per fission for release over 2~s, 1.29 for release over 20~s, and 5.21 for release over six months. The 2- and 20-s durations bracket the observed arrival of Starfish Prime debris in the conjugate atmosphere~\protect\citesupp{supp-dArcy1965}; the six-month case is an extreme upper-bound scenario representing prolonged retention of radioactive debris.

The lower and central counts are injected at $t=0$, approximating their short emission intervals relative to the three-hour belt timestep. In the upper case, 0.404 electrons per fission are injected initially and subsequent emission is added according to Eq.~\ref{eq:betacumulative}, capped at 5.207.

The electron kinetic-energy spectrum uses the Carter distribution~\protect\citesupp{supp-carter1959},
\begin{equation}
J(E)\propto\beta_v(E)\exp\!\left[-0.2938(\gamma-1)-0.0144(\gamma-1)^2\right],
\qquad \gamma=1+\frac{E}{m_ec^2},
\label{eq:carter}
\end{equation}
where $\beta_v=v/c=\sqrt{1-\gamma^{-2}}$. The spectrum is represented by 24 logarithmically spaced energies from 50~keV to 8~MeV. Discrete weights are proportional to $J(E_i)\Delta E_i$ and normalized to unity.

\subsection*{Burst region and injection shells}

The expanding weapon debris determines which field lines are available for electron injection. We follow the magnetic-confinement and equal-mass arguments of the Los Alamos Electron Source Model~\protect\citesupp{supp-cowee2012}, with an explicit IGRF field-line construction. Debris kinetic energy is $E_{\mathrm{KE}}=0.25Y$, with $Y$ expressed in joules. For a Maxwellian speed distribution with mean speed $U_d=1500$~km\,s$^{-1}$, the debris mass is
\begin{equation}
M_d=\frac{16E_{\mathrm{KE}}}{3\pi U_d^2}.
\label{eq:debrismass}
\end{equation}
For a 1-Mt detonation, this gives $M_d=789$~kg. The magnetic-confinement radius is
\begin{equation}
R_B=0.8\left(\frac{3\mu_0E_{\mathrm{KE}}}{2\pi B_b^2}\right)^{1/3},
\label{eq:rb}
\end{equation}
where $B_b$ is the field strength at the burst. The factor 0.8 is the adopted reduction from the magnetic-energy estimate.

Cross-field expansion is limited by the smaller of magnetic confinement and the distance over which the debris sweeps up its own mass in ionized air. Upward and downward equal-mass radii are calculated separately by integrating the ambient mass over the corresponding hemispheres:
\begin{equation}
\begin{aligned}
\frac{M_d}{2} &= \int_{\Omega_{\mathrm{hem}}}\int_0^{R_m}
\rho_{\mathrm{ion}}\!\left[h(s,\theta)\right]s^2\,ds\,d\Omega,\\
h(s,\theta) &= \sqrt{r_b^2+s^2+2r_bs\cos\theta}-R_E,
\end{aligned}
\label{eq:equalmass}
\end{equation}
where $r_b$ is the geocentric burst radius. The adopted ionized-air profile~\protect\citesupp{supp-cowee2012} is
\begin{equation}
\rho_{\mathrm{ion}}(h,\lambda_m)=
\frac{\rho_0e^{\ell(h)}}{\cos^2\lambda_m},
\qquad
\ell(h)=
\begin{cases}
-(h-400)/100, & h\geq320~\mathrm{km},\\
0.8-(h-320)/20, & 180\leq h<320~\mathrm{km},\\
7.8-(h-180)/10, & h<180~\mathrm{km},
\end{cases}
\label{eq:rhoion}
\end{equation}
with $\rho_0=8\times10^{-15}$~kg\,m$^{-3}$, $h$ in kilometres and magnetic latitude $\lambda_m$ capped in magnitude at $85^{\circ}$. This profile is distinct from the neutral atmosphere used for orbital drag. The horizontal equal-mass radius satisfies
\begin{equation}
M_d=\frac{4}{3}\pi R_{M,h}^{3}\rho_{\mathrm{ion}}(h_b,\lambda_{m,b}).
\label{eq:horizontalmass}
\end{equation}

The cross-field region is an asymmetric disk perpendicular to the field at the burst. Its horizontal extent is $R_h=\min(R_B,R_{M,h})$, with upward and downward extents $R_{\mathrm{up}}=\min(R_B,R_{M,\mathrm{up}})$ and $R_{\mathrm{dn}}=\min(R_B,R_{M,\mathrm{dn}})$ along the local vertical projected perpendicular to the field. For the reference burst, $R_B=944$~km and $(R_h,R_{\mathrm{up}},R_{\mathrm{dn}})=(394,708,242)$~km.

The disk is divided into area-weighted cells, and the IGRF field line through each cell determines its shell. Along each direction of a flux tube, the swept ionized-air mass per unit initial area is
\begin{equation}
m(s)=\int_0^s\rho_{\mathrm{ion}}\!\left[\mathbf{x}(s')\right]
\frac{B_{\mathrm{disk}}}{B(s')}\,ds'.
\label{eq:fluxmass}
\end{equation}
The factor $B_{\mathrm{disk}}/B$ accounts for the changing flux-tube area. Field-aligned debris expansion stops when the swept mass reaches the assigned debris mass or the atmosphere is encountered. The reference source calculation uses 1,536 cross-field cells; the location map uses 192 cells. For the reference burst, the cells reach shells $L=1.31$--1.51.

\subsection*{Artificial-belt evolution}

The trapped population is evolved on a uniform grid $1.02\leq L\leq6$ with $\Delta L=0.01$:
\begin{equation}
\frac{\partial N}{\partial t}
=\frac{\partial}{\partial L}\left(D_{LL}\frac{\partial N}{\partial L}\right)
-\frac{N}{\tau(E,L)}+Q(L,E,t).
\label{eq:diffusion}
\end{equation}
Radial diffusion uses the Brautigam--Albert magnetic coefficient~\protect\citesupp{supp-brautigam2000},
\begin{equation}
D_{LL}=10^{0.506K_p-9.325}L^{10}~\mathrm{day}^{-1},
\qquad K_p=2.
\label{eq:dll}
\end{equation}
Equation~\ref{eq:diffusion} is solved by backward Euler with a tridiagonal solver, three-hour timesteps and zero-flux boundaries. The calculation uses this simplified radial operator rather than a full phase-space-density diffusion equation. At $K_p=2$, $D_{LL}$ is approximately $3\times10^{-8}$~day$^{-1}$ at $L=1.2$ and $5\times10^{-6}$~day$^{-1}$ at $L=2$, corresponding to characteristic diffusion distances $\sqrt{2D_{LL}t}$ of approximately 0.002--0.02 over 60 days.

Electron losses use the empirical parameterization
\begin{align}
\tau_0(L)&=15+1485\exp\!\left[-\left(\frac{L-1.4}{0.55}\right)^2\right]
+\frac{45}{1+\exp[-(L-3.5)/0.4]},\\
a(L)&=\mathrm{clip}\!\left(\frac{L-1.08}{0.10},0,1\right),\\
\tau(E,L)&=\left[\tau_0(L)a(L)+1-a(L)\right]
\left(\frac{\max(E,0.1)}{0.5}\right)^{0.3},
\label{eq:electronlifetime}
\end{align}
where energy is in MeV and lifetimes are in days, bounded between 0.5 days and eight years. This prescribed form represents long lifetimes in the inner belt, shorter lifetimes in the slot region and atmospheric loss on low shells.

Electron densities use dipole shell volumes above 80~km,
\begin{equation}
\begin{aligned}
V_i &= 2\pi R_E^3\int_{L_i^-}^{L_i^+}\int_{-\lambda_m}^{\lambda_m}
L^2\cos^7\lambda\,d\lambda\,dL,\\
\cos^2\lambda_m &= \frac{R_E+80~\mathrm{km}}{LR_E}.
\end{aligned}
\label{eq:shellvolume}
\end{equation}
This volume approximation is separate from the IGRF trapping calculation and its 120-km loss boundary. The belt evolution is a reduced-order reconstruction rather than a data-assimilative radiation-belt calculation~\protect\citesupp{supp-reeves2012,supp-lee2024}.

\subsection*{Object-dependent collision geometry}

Collision areas are calculated from orientation-averaged geometric proxies, independently of the aerodynamic areas used for drag. Starlink generations use bus-and-array envelopes of $9\times2.8\times0.2$~m for v1.x and $29\times4.1\times0.3$~m for V2 families~\protect\citesupp{supp-spacex2022geometry}. Using the same V2 envelope for optimized and direct-to-cell variants is an estimate. Other objects use GCAT dimensions with box, cylinder or sphere proxies~\protect\citesupp{supp-gcat2026}. Missing dimensions receive a sphere with the median area-equivalent radius of the corresponding payload or rocket-body population; assignments and catalogue flags are retained in the geometry ledger.

For independently isotropic orientations of convex proxies, the mean pair area is
\begin{equation}
\bar\sigma_{ab}=\bar A_a+\bar A_b+\frac{\bar P_a\bar P_b}{2\pi},
\label{eq:geometricarea}
\end{equation}
where $\bar A$ and $\bar P$ are mean projected area and perimeter. This follows the projected collision-region construction used for extended objects~\protect\citesupp{supp-garciapelayo2018}. For a box of sides $x,y,z$, $\bar A=(xy+xz+yz)/2$ and $\bar P=\pi(x+y+z)/2$. For a fragment, we use its breakup-model area and the perimeter of an equal-area circle. These surrogates include gaps within solar-array envelopes and are not detailed spacecraft silhouettes.

Confirmed docked ISS and Tiangong entries are grouped into two collision identities; independent satellites with identical launch-group elements remain separate. Station geometry is the sum of component collision regions. In Eq.~\ref{eq:geometricarea}, the first two terms become $K_b\bar A_a+K_a\bar A_b$, where $K$ counts geometric components. Overlapping component regions are not subtracted. Long antennas and tethers use a body plus a thin-wire contribution to projected perimeter, rather than a filled envelope spanning their length. Their retained deployed lengths are uncertain. Separate sensitivity runs vary these special-object dimensions.

For comparison, we use the encounter-distance prescription of
Thiele and colleagues~\protect\citesupp{supp-thiele2026}: 10~m for intact--intact
pairs, 5~m for intact--debris pairs and 10~cm for debris--debris
pairs, denoted here as 10-5-10. These correspond to approximate
pair areas of 300, 79 and 0.03~m$^2$, respectively. We also test
their smaller-distance prescription, using 4.8~m, 2.4~m and
10~cm. These prescriptions describe close encounters rather than
object-dependent physical impacts. In the coupled sensitivity
runs, we treat each modeled encounter as a collision and use
the assigned areas consistently for rates and fragment production.

As an implementation check, applying these prescriptions to the
June catalogue with 1-km shells gives clocks of approximately
5.5 and 22.8 days, reproducing the approximately 5.5- and 23-day
values reported by Thiele and colleagues~\protect\citesupp{supp-thiele2026}.
Using our adopted 10-km shells gives approximately 5.8 and
24.2 days.

\subsection*{Collision rates and intact-object accounting}

Radiation failures are coupled to an altitude-resolved collision model adapted from the CRASH Clock formalism of Thiele and colleagues~\protect\citesupp{supp-thiele2026} and the NASA breakup treatment used by Thiele and Boley~\protect\citesupp{supp-thiele2022,supp-johnson2001}. The central calculation uses 1.29 emitted electrons per fission and the uniform 40~krad(Si) threshold. It estimates shell-averaged physical-impact rates using geometric approximations; it does not search for individual conjunctions. Integrating these rates gives expected collision counts, rather than a randomly sampled sequence of individual events. Unless otherwise stated, totals include all pair types and both catastrophic and cratering outcomes. The Poisson expression below converts the integrated rate into the probability of at least one collision.

Operational payloads, radiation-disabled payloads, payloads without radiation histories, rocket bodies, pre-existing catalogued debris and new collision fragments remain distinct. An intact-object ledger records each identity's surviving fraction, collision destruction, mass loss and atmospheric removal. ``Intact'' identifies a retained spacecraft or rocket body, including a target that survives a cratering event.

For unordered classes or fragment bins $a,b$ in shell $h$, the rate is
The pair-rate equation is given in Methods (Eq.~\ref{eq:pairrate}).
where $V_h$ is the shell volume and $s_{ab}$ is one half for identical classes and one otherwise. For different classes, $\mathcal{G}_{ab,h}=N_{a,h}N_{b,h}\bar\sigma_{ab,h}$. For an identical intact class, we subtract $\sum_i u_i^2w_{ih}^2\sigma_{ii}$ from this product to exclude an object encountering itself; $u_i$ is its survival fraction and $w_{ih}$ its orbital residence fraction. Fragment bins use the continuum population product. The relative speed assumes isotropic velocity directions at the local circular speed. We use 10-km shells from 0 to 2000~km.

The residual-risk factor is $p_{ab}=0.01$ when at least one partner is operational and one otherwise. Avoidance credit is applied once per pair. A disabled spacecraft can therefore benefit from an operational partner's maneuver. Applying this benefit to all tracked fragments is an optimistic assumption; detectability and individual maneuver constraints are not modeled.

\subsection*{Intact masses, areas and atmospheric drag}

Starlink generations are assigned using NORAD identifiers, spacecraft metadata and launch groups from McDowell's GCAT snapshot~\protect\citesupp{supp-gcat2026}. Generation-specific masses follow public technical information~\protect\citesupp{supp-sutton2025,supp-spacex2024,supp-gcat2026}; area-to-mass ratios represent uncontrolled orientations (Supplementary Table~\ref{tab:intact_properties}). The v1.0 ratio follows an orientation-averaged thin-panel estimate using the bus and array geometry discussed by Fitzpatrick et al.~\protect\citesupp{supp-fitzpatrick2026}. Other payloads and rocket bodies use available GCAT dry masses, with recorded 260- and 1000-kg fallbacks, respectively. Non-Starlink masses are separated into five ranges for collision classification. Docked masses are counted once; the ISS uses a rounded 420-tonne nominal mass proxy~\protect\citesupp{supp-nasaISS}, and Tiangong uses the summed component dry masses. These are representative masses, not measurements of each spacecraft's current fuel and configuration.

For non-Starlink payloads we retain $A/m=0.010$~m$^2$\,kg$^{-1}$ from an ESA reference case~\protect\citesupp{supp-esa2025}; rocket bodies use the assumed 0.005~m$^2$\,kg$^{-1}$. Docked complexes use their projected-area proxy for drag. We use $C_D=2.2$, ESA's default for long-duration lifetime calculations when a better estimate is unavailable~\protect\citesupp{supp-esa2025}. Aerodynamic area remains fixed for each identity; cratering can reduce mass and increase $A/m$.

Operational payloads maintain their orbits until radiation failure; other intact objects undergo drag from the start. Orbit-averaged drag updates perigee and eccentricity, after which Keplerian residence fractions are recalculated~\protect\citesupp{supp-fitzpatrick}. Shell heights use a 6371-km spherical reference and drag uses 6378~km; heights are converted through physical radius before density lookup. Any surviving fraction is removed when drag-reference perigee reaches 200~km. Initial perigees and eccentricities are calculated from SGP4 Cartesian positions and velocities at the burst epoch.

\subsection*{Fragmentation and mass conservation}

At each step, expected collisions $R_{ab,h}\Delta t$ remove colliders or erode surviving targets. Following the NASA Standard Breakup Model~\protect\citesupp{supp-johnson2001}, an impact at the representative speed $v=10$~km\,s$^{-1}$ is catastrophic when $M_pv^2/(2M_t)\geq40$~kJ\,kg$^{-1}$, where $M_t\geq M_p$. Both objects are then removed. Otherwise the projectile is removed, while the target retains its identity with reduced mass. Ejected mass is
\begin{equation}
M_e=\begin{cases}
M_t+M_p,&\text{catastrophic},\\
\min\{M_t+M_p,M_p[v/(1~\mathrm{km\,s^{-1}})]^2\},&\text{cratering}.
\end{cases}
\label{eq:ejectedmass}
\end{equation}
The target loses $M_e-M_p$ in a cratering event; additional functional failure from that impact is not modeled. Outcomes use the current mean mass within each class and shell. Losses are assigned to intact identities by orbital residence. Collision demand is capped by the available population, including consumption of two objects for an identical-pair catastrophic event.

Proposed fragment numbers follow~\protect\citesupp{supp-johnson2001,supp-thiele2022}
\begin{equation}
N_f(>L_c)=0.1M_e^{0.75}L_c^{-1.71},\qquad
A(L_c)=0.556945L_c^{2.0047077},\qquad m=A/\alpha,
\label{eq:fragmentmass}
\end{equation}
with SI units and area-to-mass ratio $\alpha$. The sampled spectrum covers 0.1--1~m, using the payload-family breakup distribution with $10^{-3}\leq\alpha\leq10$~m$^2$\,kg$^{-1}$. Samples are compressed into 60 joint size--area-to-mass cells retaining number, area and mass moments, plus eight bins for pieces carrying residual breakup mass. The same distribution initializes background debris because individual catalogue sizes and masses are unavailable.

We limit production to the available ejected mass and assign any remainder to explicit mass-carrying pieces in the largest size range. Material from 1~mm to 10~cm enters a separate mass ledger and is excluded from subsequent rates. It is not counted as atmospheric removal. Number and mass are carried together in each orbital bin. Mean mass can change through injection, collisions and mixing. Mass accounting balances surviving intact objects, debris, atmospheric removal, escaping fragments and sub-10-cm material.

\subsection*{Fragment velocity kicks and orbital transport}

New fragments receive isotropically directed velocity kicks drawn from the NASA Standard Breakup Model distribution~\protect\citesupp{supp-johnson2001}, as used by Thiele and Boley~\protect\citesupp{supp-thiele2022}:
\begin{equation}
\begin{aligned}
\log_{10}\!\left[\Delta v/(\mathrm{m\,s^{-1}})\right]
&=0.9\log_{10}\!\left[\alpha/(\mathrm{m^2\,kg^{-1}})\right]\\
&\quad+2.9+0.4Z,
\end{aligned}
\label{eq:fragmentkick}
\end{equation}
where $Z$ is drawn from a normal distribution with mean zero and unit variance. We sample 128 velocity kicks for each representative area-to-mass ratio and add them to the circular-orbit velocity at the collision altitude. This approximates the motion of the colliding objects rather than reconstructing their actual center-of-mass velocity. The kick distributions are calculated at 16 area-to-mass ratios between $10^{-4}$ and 10~m$^2$\,kg$^{-1}$ and interpolated between these values. Outside this range, we use the nearest calculated distribution for the kicks while retaining the fragment's actual area-to-mass ratio for drag. Fragments that escape Earth or have a perigee at or below 200~km are accounted for separately.

We follow the remaining fragments by their perigee and eccentricity, using 20-km perigee intervals and 17 eccentricity values while conserving fragment number and mass. Each fragment population contributes to an altitude band according to the fraction of its orbit spent there. Breakup kicks can therefore spread debris both above and below the collision altitude. Pre-existing debris starts on its catalogued January orbit without an additional kick. Atmospheric drag gradually lowers both populations until their perigees reach 200~km, when they are removed. The calculation follows changes in orbital altitude and shape, but does not resolve inclination, position along the orbit or orbital precession caused by Earth's nonspherical gravity.

\subsection*{Atmospheric forcing and numerical integration}

Atmospheric densities use the low-, moderate- and high-activity CIRA-2012 profiles~\protect\citesupp{supp-cira2012}, corresponding to F10.7 solar radio fluxes of 65, 140 and 250 solar flux units. The central calculation uses the NASA/MSFC October 2026 release: smoothed observations, median forecast and mean-cycle extension~\protect\citesupp{supp-nasa2026}. Monthly fluxes are interpolated in time and neighboring profiles blended linearly. Density is interpolated logarithmically with altitude and extrapolated above 900~km. This does not resolve storms, atmospheric rotation or geographic variations, or independently vary geomagnetic activity.

We compare the 5th and 95th solar-flux paths, Starlink drag-property bounds, fixed CIRA profiles, and the stated geometric alternatives. The matched no-detonation run uses identical initial data and forcing; operational objects maintain their orbits, while other objects and debris evolve. Neither run includes new launches or scheduled disposals.

The coupled model uses three-hour steps through day 60 and daily steps thereafter. Radiation-failure states are fixed after day 60; dose is not recomputed along descending orbits. Each intact object's surviving, collision-destroyed and removed fractions sum to one. Two-year timestep, orbital-grid, kick-sampling, breakup-closure and special-dimension tests change collision counts by less than 0.5\% and new-fragment populations by less than 3\%. These assess numerical sensitivity, not physical validation. Ten-year mass residuals remain below $10^{-12}$ of initial mass. Detailed diagnostics accompany the outputs; drag and residence fractions are checked against direct propagation and dense orbital sampling.

\subsection*{Collision hazard for an inserted satellite}

Supplementary Fig.~\ref{fig:orbit_collision_rate_relative_to_background} considers one additional payload on a circular orbit, with a $9\times2.8\times0.2$-m isotropic box proxy. It does not alter the populations. Its instantaneous rate is
\begin{equation}
\lambda_{\mathrm{test}}(t)=\sum_h w_h\frac{v_h}{V_h}
\sum_b p_{\mathrm{test},b}\bar\sigma_{\mathrm{test},b}N_{b,h}(t),
\label{eq:tracerrate}
\end{equation}
with no identical-pair half factor. For a non-maneuvering test object, avoidance credit applies only against operational partners; for a maneuvering object it applies against every modeled partner. We plot $\lambda_{\mathrm{detonation}}/\lambda_{\mathrm{control}}$ for identical test geometry and matched environments. Cells without a positive control rate remain undefined.

The experiment resolves altitude, not inclination or conjunction phase. It assumes the test satellite survives the radiation environment and excludes impacts from debris below 10~cm. Its relative rates therefore do not determine full survivability or a safe date for renewed operations.

\begingroup
\footnotesize\singlespacing
\bibliographystylesupp{unsrtnat}
\bibliographysupp{references-supp}
\endgroup
\endgroup

\end{document}